\documentclass[%
 reprint,
 amsmath,amssymb,
 aps,
]{revtex4-2}

\usepackage{graphicx}% Include figure files
\usepackage{dcolumn}% Align table columns on decimal point
\usepackage{bm}% bold math
 \usepackage{lineno}
\usepackage{multirow}
\usepackage{tabularx}
\usepackage{mathtools}
\usepackage{slashed}
\usepackage{booktabs}
\usepackage{url}

\usepackage[maxfloats=100]{morefloats}
\usepackage{color,xcolor}
\usepackage[normalem]{ulem}
\usepackage{subfig}

\newcommand{\PZ}{\ensuremath{\mathrm{Z}}}
\newcommand{\PH}{\ensuremath{\mathrm{H}}}
\newcommand{\PW}{\ensuremath{\mathrm{W}}}
\newcommand{\MGMCatNLO}{MadGraph5\_aMC@NLO} 

\newcommand{\delphes}{{\sc Delphes}}

\newcommand{\met}{\ensuremath{\slashed{E}_T}}

\newcommand{\abinv}{\mbox{\ensuremath{\mathrm{ab^{-1}}}}}

\newcommand{\PTeV}{\ensuremath{\mathrm{TeV}}}
\newcommand{\PGeV}{\ensuremath{\mathrm{GeV}}}

\begin{document}
\def\wng{{{\cal W}}_0^\gamma}
\def\wnz{{{\cal W}}_0^Z}
\def\wcg{{{\cal W}}_c^\gamma}
\def\wcz{{{\cal W}}_c^Z}
\def\wuz{{{\cal W}}_1^Z}
\def\wdz{{{\cal W}}_2^Z}
\def\wtz{{{\cal W}}_3^Z}
\def\zng{{{\cal Z}}_0^\gamma}
\def\zcg{{{\cal Z}}_c^\gamma}
\def\znz{{{\cal Z}}_0^Z}
\def\zcz{{{\cal Z}}_c^Z}

%\preprint{APS/123-QED}

\title{Searches for multi-Z boson productions and anomalous gauge boson couplings at a muon collider}% Force line breaks with \\

\author{Ruobing Jiang}
 \altaffiliation{School of Physics and State Key Laboratory of Nuclear Physics and Technology, Peking University.} 
 
\author{Chuqiao Jiang}%
 \email{chuqiao.jiang@cern.ch }
\affiliation{School of Physics and State Key Laboratory of Nuclear Physics and Technology, Peking University}%

\author{Alim Ruzi}%
\affiliation{School of Physics and State Key Laboratory of Nuclear Physics and Technology, Peking University}%

\author{Tianyi Yang}%
\affiliation{School of Physics and State Key Laboratory of Nuclear Physics and Technology, Peking University}%

\author{Yong Ban}%
\affiliation{School of Physics and State Key Laboratory of Nuclear Physics and Technology, Peking University}%

\author{Qiang Li}%
\affiliation{School of Physics and State Key Laboratory of Nuclear Physics and Technology, Peking University}%

\begin{abstract}
Multi-boson productions can be exploited as novel probes either for standard model precision tests or new physics searches, and have become one of those popular topics in the ongoing LHC experiments, and in future collider studies, including those for electron–positron and muon–muon colliders. Here we focus on two examples, i.e., $\PZ\PZ\PZ$ direct productions through $\mu^+\mu^-$ annihilation at a $1\,\PTeV$ muon collider, and $\PZ\PZ$ productions through vector boson scattering(VBS) at a $10\,\PTeV$ muon collider, with an integrated luminosity of $10\,\abinv$. Various channels are considered, including, such as $\PZ\PZ\PZ \rightarrow 4\ell2\nu$ and $\PZ\PZ\PZ \rightarrow 4\ell$+2jets, etc. Expected significance on these multi-Z boson production processes are provided based on a detailed Monte Carlo study and signal background analysis. Sensitives on anomalous gauge boson couplings are also presented.
\end{abstract} 
\maketitle

\section{\label{sec:level1}INTRODUCTION}
The Standard Model (SM) is based on $SU(3)_{C}\otimes SU(2)_{L}\otimes U(1)_{Y}$ gauge symmetry group and describes the interactions among all elementary particles~\cite{Green:2016trm}. In 2012, The discovery of Higgs boson by the CMS and ATLAS experiments~\cite{ATLAS:2012yve,CMS:2012qbp} at the Large Hadron Collider (LHC) marked a great success of the SM physics. The High-Luminosity LHC (HL-LHC), together with other future colliders, such as muon colliders, will not only enable people to make more precise measurements on characterising properties of the SM physics, but also to unravel the undiscovered phenomena lying beyond the SM physics.

Recently, a muon collider working at a centre of mass (COM) energy of TeV scale has received revived interest from the community of high-energy physics \cite{Roser:2022sht,Long:2020wfp}. As muons are approximately 200 times heavier than electrons, energy loss caused by synchrotron radiation for muons is much less than for electrons. Moreover, muon-muon collisions provide cleaner environment than proton-proton collisions. These features make a muon collider an attractive energy efficient machine to explore high-energy physics. 

A muon collider offers numerous opportunities to study elementary particle physics\cite{MuonCollider:2022xlm, Accettura:2023ked}. As one of the scenarios, when the COM energy is around $1\,\PTeV$, $\mu^{+}\mu^{-}$ annihilation acts as the dominant production mechanism. At multi-TeV scale, muons have a high probability to emit electroweak (EW) radiation, thus a high energy muon collider can also serve as a vector boson collider. Both collision modes present spectacular playground for both the search for the origin of EW symmetry breaking (EWSB) and for the EW interactions Beyond Standard Model (BSM), such as anomalous gauge boson interactions~\cite{Amarkhail:2023xsc,Yang:2022fhw,Yang:2020rjt,Dong:2023nir,Zhang:2023yfg,Jahedi:2022duc,Jahedi:2023myu}.  

At the current and future colliders, multiboson production is an interesting topic sensitive to the non-abelian character of the SM~\cite{Green:2016trm,Langacker:1994amo}. In particular, the presence of anomalous quartic gauge boson interactions~\cite{Eboli:2003nq,Degrande:2012wf,Degrande:2013kka} can be probed through tri-boson production, and di-boson production through vector boson scattering. There has been a lot of researches on this topic at the LHC~\cite{CMS:2020hjs,Kunkle:2015swy}. In this paper, we focus on two examples, i.e., $\PZ\PZ\PZ$ direct productions through $\mu^{+}\mu^{-}$ annihilation at a $1\,\PTeV$ muon collider, and $\PZ\PZ$ productions through VBS process at a $10\,\PTeV$ muon collider, with an integrated luminosity of $10\,\abinv$. 

\section{\label{sec:level2} Multiboson and Anomalous Quartic Gauge Couplings}

Precision measurements of multiboson production allow a basic test of SM, and provide a model independent method to search for BSM at the $\PTeV$ scale~\cite{CMS:2020hjs}. In our study, we focus on $\PZ\PZ\PZ$ direct productions, and $\PZ\PZ$ productions through vector boson scattering. Both processes are sensitive to non-abelian gauge boson interactions and the structure of EW symmetry breaking. These multiboson processes represent an important avenue to test anomalous triple gauge couplings (aTGCs) and anomalous quartic gauge couplings (aQGCs)~\cite{Green:2016trm}, and to search for possible modification of these vertices from new physics~\cite{Kunkle:2015swy}.

%The SM Lagrangian is :
%\begin{equation}   
%\mathcal{L}_{EW}=\mathcal{L}_{K}+\mathcal{L}_{N}+\mathcal{L}_{C}+\mathcal{L}_{H}+\mathcal{L}_{HV}+\mathcal{L}_{VVV}+\mathcal{L}_{VVVV}+\mathcal{L}_{Y}.
%\end{equation}

 Anomalous modifications of gauge couplings can be parameterized through Effective Field Theory (EFT) adding higher order modifications to the SM Lagrangian:
\begin{equation}
\mathcal{L}^{NP}=\mathcal{L}^{4(SM)}+\frac{1}{\Lambda}\mathcal{L}^{5}+\frac{1}{\Lambda^2}\mathcal{L}^{6}+\frac{1}{\Lambda^3}\mathcal{L}^{7}+\frac{1}{\Lambda^4}\mathcal{L}^{8}+...
\end{equation}

The higher order terms are suppressed by a mass scale $\Lambda$, representing the scale of new physics beyond the SM. The odd dimensions terms are not considered because they will not influence multiboson production measurements. The dimension-6 operators are related to aTGCs and the dimension-8 operators are related to aQGCs. 

Notice aQGCs can be realized by introducing some new heavy bosons, which contribute to aQGCs at tree-level, while one-loop suppressed in aTGCs \cite{Belanger:1992qh,Arzt:1994gp,Eboli:2003nq}. Furthermore, as aTGCs are currently tested to be in good agreement with the SM via many experimental studies, our study thus mainly focuses on genuine aQGCs.

To express the aQGC contributions model-independently, an effective field theory of the EW breaking sector~\cite{Belyaev:1998ih,Eboli:2000ad,Eboli:2003nq,Eboli:2006wa,Belanger:1999aw,Eboli:2016kko} is utilized. When the $SU(2)_{L}\otimes U(1)_{Y}$ is represented linearly, the lowest order genuine aQGC operators parameterized in the EFT are dimension-8 (dim-8)\cite{Eboli:2006wa, Eboli:2016kko,Degrande:2013rea,Baak:2013fwa}. The genuine aQGC operators can be expressed as~\cite{Almeida:2020ylr}:
\begin{equation}
\begin{array}{lc}
\vspace{1em}
  {\cal O}_{S,0} = 
\left [ \left ( D_\mu \Phi \right)^\dagger
 D_\nu \Phi \right ] \times 
\left [ \left ( D^\mu \Phi \right)^\dagger
D^\nu \Phi \right ]
&,
\\   \vspace{1em}
  {\cal O}_{S,1} =
 \left [ \left ( D_\mu \Phi \right)^\dagger
 D^\mu \Phi  \right ] \times
\left [ \left ( D_\nu \Phi \right)^\dagger
D^\nu \Phi \right ]
&,
\\   \vspace{1em}
  {\cal O}_{S,2} =
 \left [ \left ( D_\mu \Phi \right)^\dagger
 D_\nu \Phi  \right ] \times
\left [ \left ( D^\nu \Phi \right)^\dagger
D^\mu \Phi \right ]
&,
\\   \vspace{1em}
%%%%%%%%%%%%%%%%%%%%%%%%%%%%
 {\cal O}_{M,0} =   \hbox{Tr}\left [ \widehat{W}_{\mu\nu} \widehat{W}^{\mu\nu} \right ]
\times  \left [ \left ( D_\beta \Phi \right)^\dagger
D^\beta \Phi \right ]
&,
\\   \vspace{1em}
 {\cal O}_{M,1} 
=   \hbox{Tr}\left [ \widehat{W}_{\mu\nu} \widehat{W}^{\nu\beta} \right ]
\times  \left [ \left ( D_\beta \Phi \right)^\dagger
D^\mu \Phi \right ]
&,
\\   \vspace{1em}
 {\cal O}_{M,2} =   \left [ B_{\mu\nu} B^{\mu\nu} \right ]
\times  \left [ \left ( D_\beta \Phi \right)^\dagger
D^\beta \Phi \right ]
&,
\\   \vspace{1em}
 {\cal O}_{M,3} =   \left [ B_{\mu\nu} B^{\nu\beta} \right ]
\times  \left [ \left ( D_\beta \Phi \right)^\dagger
D^\mu \Phi \right ]
&,
\\   \vspace{1em}
  {\cal O}_{M,4} = \left [ \left ( D_\mu \Phi \right)^\dagger \widehat{W}_{\beta\nu}
 D^\mu \Phi  \right ] \times B^{\beta\nu}
&,
\\   \vspace{1em}
  {\cal O}_{M,5} = \left [ \left ( D_\mu \Phi \right)^\dagger \widehat{W}_{\beta\nu}
 D^\nu \Phi  \right ] \times B^{\beta\mu}+ {\rm h.c.}
&,
\\   \vspace{1em}
%   {\cal O}_{M,6} = \left [ \left ( D_\mu \Phi \right)^\dagger \widehat{W}_{\beta\nu}
% \widehat{W}^{\beta\nu} D^\mu \Phi  \right ] 
% &,&
  {\cal O}_{M,7} = \left [ \left ( D_\mu \Phi \right)^\dagger \widehat{W}_{\beta\nu}
\widehat{W}^{\beta\mu} D^\nu \Phi  \right ]  
&,
\\   \vspace{1em}
%%%%%%%%%%%%%%%%%%%%%%%%%%%%%%%%%
 {\cal O}_{T,0} =   \hbox{Tr}\left [ \widehat{W}_{\mu\nu} \widehat{W}^{\mu\nu} \right ]
\times   \hbox{Tr}\left [ \widehat{W}_{\alpha\beta} \widehat{W}^{\alpha\beta} \right ]
&,
\\   \vspace{1em}
 {\cal O}_{T,1} =   \hbox{Tr}\left [ \widehat{W}_{\alpha\nu} \widehat{W}^{\mu\beta} \right ]
\times   \hbox{Tr}\left [ \widehat{W}_{\mu\beta} \widehat{W}^{\alpha\nu} \right ]
&,
\\   \vspace{1em}
 {\cal O}_{T,2} =   \hbox{Tr}\left [ \widehat{W}_{\alpha\mu} \widehat{W}^{\mu\beta} \right ]
\times   \hbox{Tr}\left [ \widehat{W}_{\beta\nu} \widehat{W}^{\nu\alpha} \right ]
&,
\\   \vspace{1em}
 {\cal O}_{T,3} =  \hbox{Tr}\left [ \widehat{W}_{\mu\nu} \widehat{W}_{\alpha\beta} \right ]
\times   \hbox{Tr}\left [ \widehat{W}^{\alpha\nu} \widehat{W}^{\mu\beta} \right ]
&,
\\   \vspace{1em}
 {\cal O}_{T,4} =   \hbox{Tr}\left [ \widehat{W}_{\mu\nu} \widehat{W}_{\alpha\beta} \right ]
\times   B^{\alpha\nu} B^{\mu\beta}
&,
\\   \vspace{1em}
 {\cal O}_{T,5} =   \hbox{Tr}\left [ \widehat{W}_{\mu\nu} \widehat{W}^{\mu\nu} \right ]
\times   B_{\alpha\beta} B^{\alpha\beta}
&,
\\   \vspace{1em}
 {\cal O}_{T,6} =   \hbox{Tr}\left [ \widehat{W}_{\alpha\nu} \widehat{W}^{\mu\beta} \right ]
\times   B_{\mu\beta} B^{\alpha\nu} 
&,
\\   \vspace{1em}
 {\cal O}_{T,7} =   \hbox{Tr}\left [ \widehat{W}_{\alpha\mu} \widehat{W}^{\mu\beta} \right ]
\times   B_{\beta\nu} B^{\nu\alpha} 
&,
\\   \vspace{1em}
 {\cal O}_{T,8} =   B_{\mu\nu} B^{\mu\nu}  B_{\alpha\beta} B^{\alpha\beta}
&,
\\   \vspace{1em}
 {\cal O}_{T,9} =  B_{\alpha\mu} B^{\mu\beta}   B_{\beta\nu} B^{\nu\alpha} 
&.
\end{array}
\label{eq:operators}
\end{equation}
where $\Phi$ stands for the Higgs doublet, the covariant derivative is given by $D_\mu \Phi = (\partial_\mu + i g W^j_\mu \frac{\sigma^j}{2} + i g^\prime B_\mu \frac{1}{2}) \Phi$ and $\sigma^j$ ($j=1,2,3$) represent the Pauli matrices. $\widehat{W}_{\mu\nu} \equiv  W^j_{\mu\nu} \frac{\sigma^j}{2}$
is the $SU(2)_L$ field strength while $B_{\mu\nu}$ stands for the $U(1)_Y$ one. %$D_\mu$ is the covariant derivative.
The effective Lagrangian with the contributions from genuine aQGC operators can be expressed as:
\begin{equation}
\begin{array}{ll}
\vspace{1em}
\mathcal{L}_{eff}
&=\mathcal{L}_{SM}+\mathcal{L}_{anomalous} \\ \vspace{1em}
&=\mathcal{L}_{SM}+\sum_{d>4} \sum_{i} \frac{f_{i}^{(d)}}{\Lambda^{d-4}} 
O_{i}^{(d)} \\ \vspace{1em}
&=\mathcal{L}_{SM}+\sum_{i}[\frac{f_{i}^{(6)}}{\Lambda^{2}}O_{i}^{(6)}]+\sum_{j}[\frac{f_{j}^{(8)}}{\Lambda^{4}}O_{i}^{(8)}]+...,
\label{eq:lagrangian}
\end{array}
\end{equation}
where $\Lambda$ is the characteristic scale and $f_{j}^{(8)}/\Lambda^{4}=f_{S, j}/\Lambda^{4}, f_{M, j}/\Lambda^{4}, f_{T, j}/\Lambda^{4}$ represent the coefficients of the corresponding aQGC operators~\cite{Eboli:2016kko}. These coefficients are expected to be zero in the SM prediction. 

\begin{figure}
\centering
\subfloat[\label{fig:a}]{
\includegraphics[width=4.2cm]{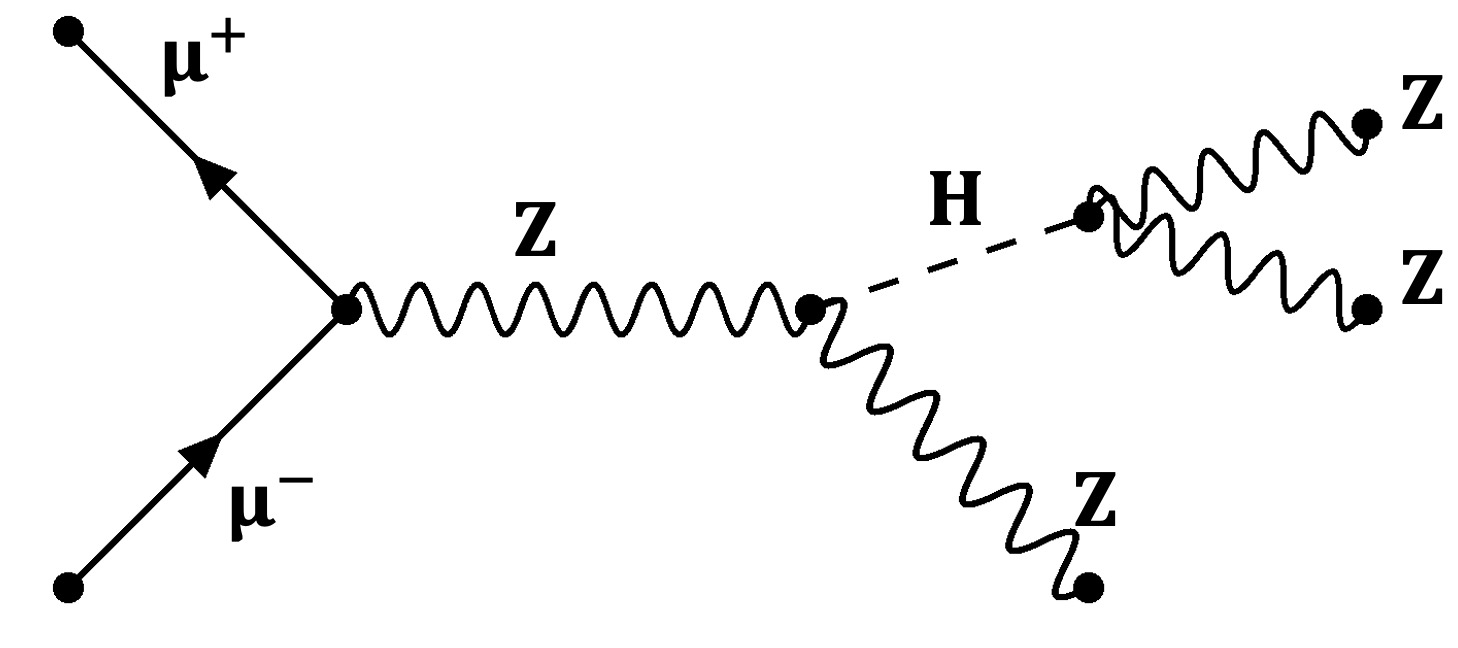}}
\subfloat[\label{fig:b}]{
\includegraphics[width=4.2cm]{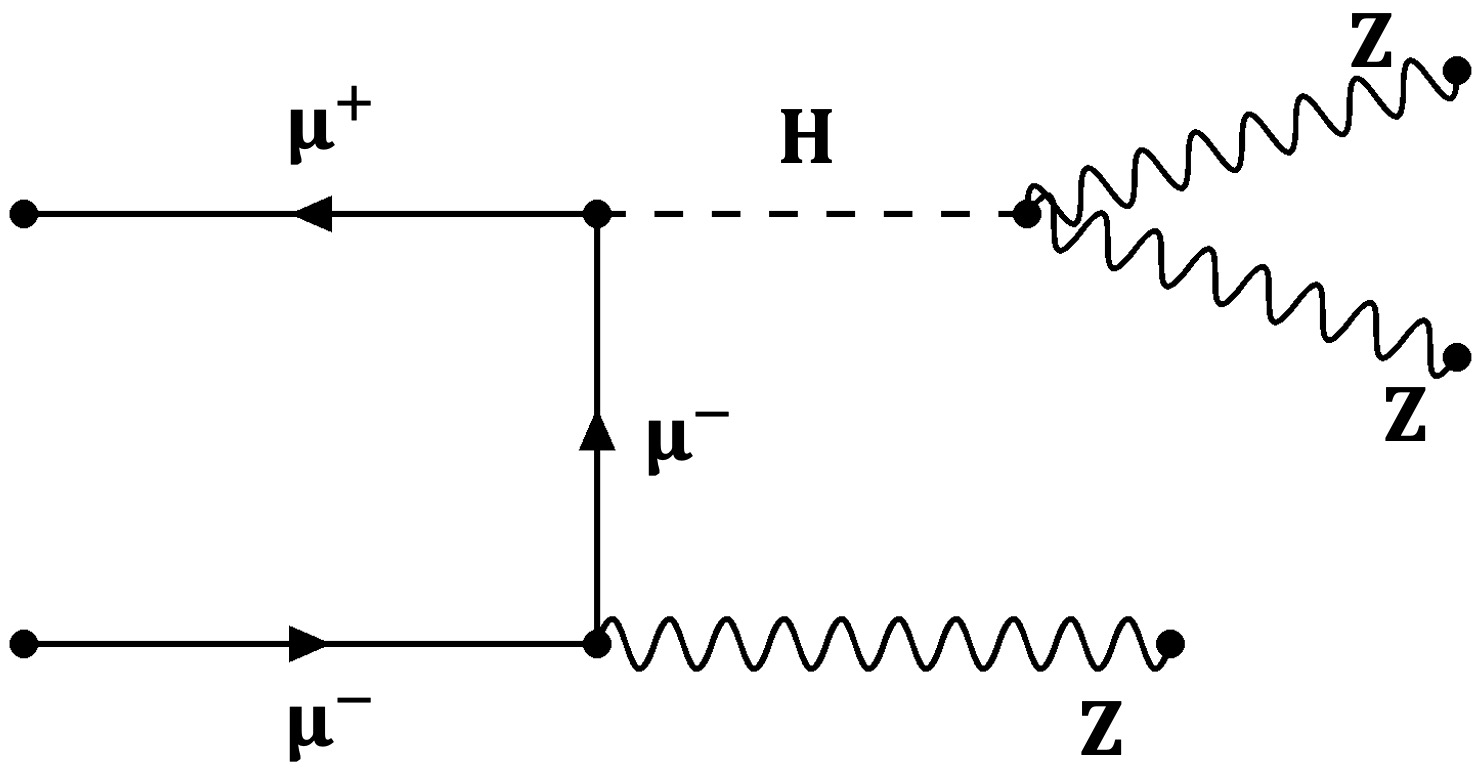}}
\\
\subfloat[\label{fig:c}]{
\includegraphics[width=4.2cm]{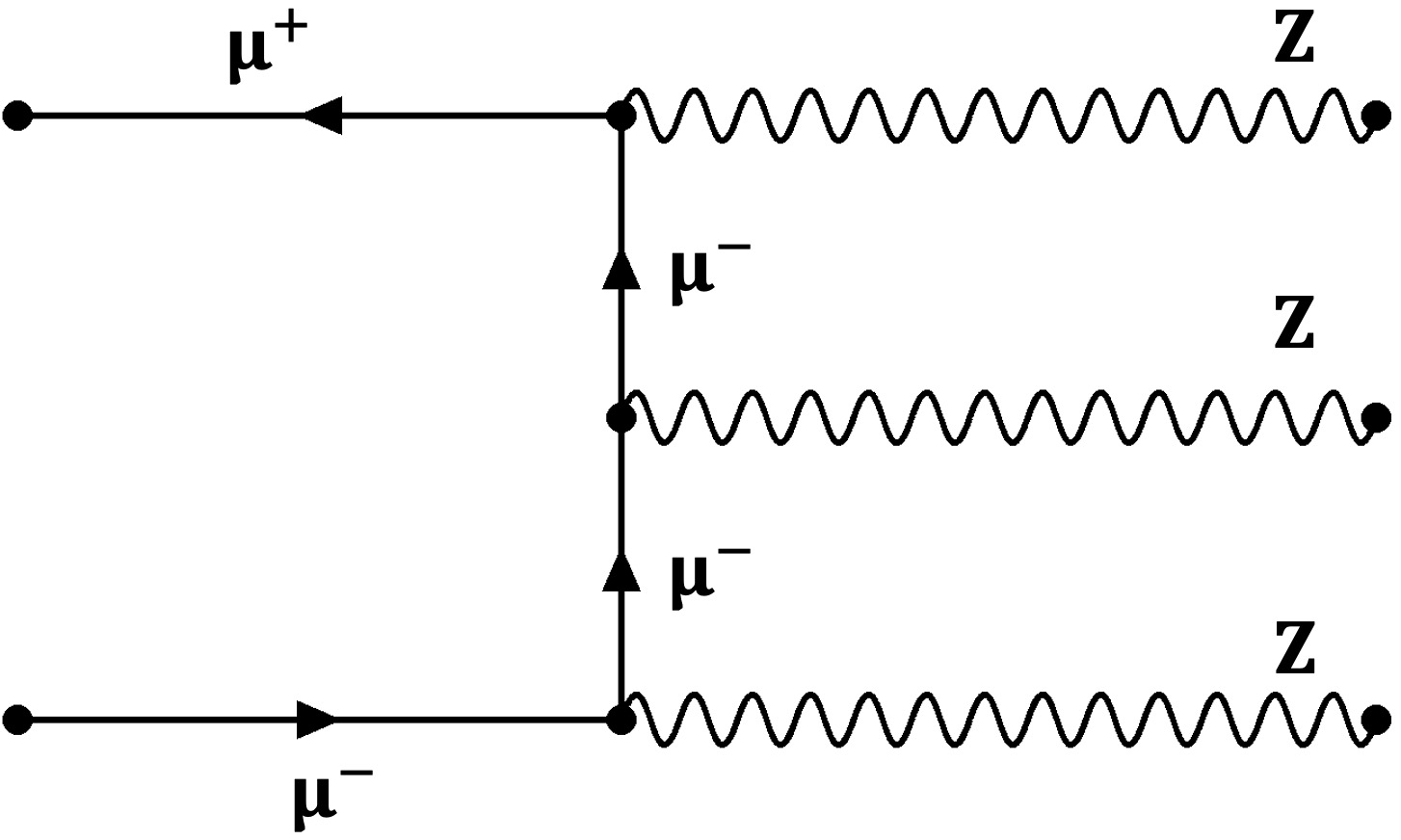}}
\subfloat[\label{fig:d}]{
\includegraphics[width=4.2cm]{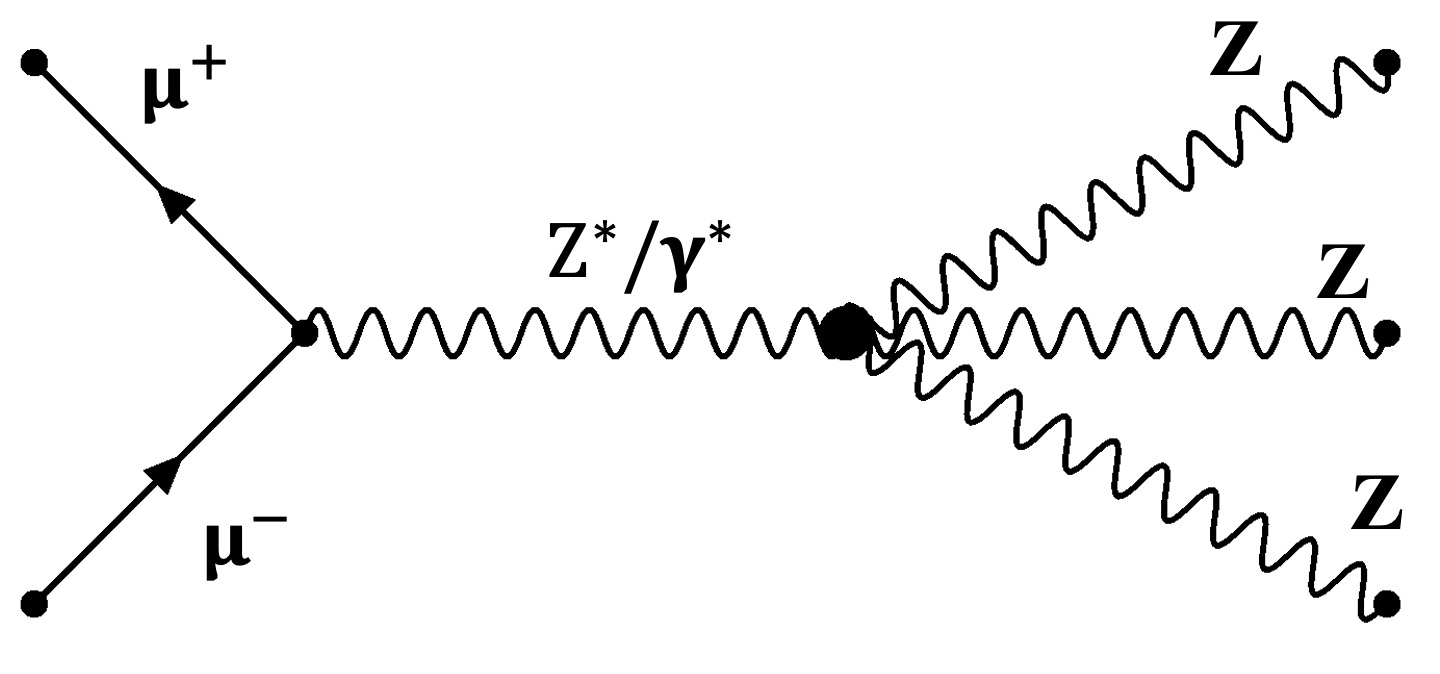}}
\caption{Example Feynman diagrams of $\PZ\PZ\PZ$ production processes at a muon collider: (a-c) are from the SM, and (d) involves quartic gauge couplings.}
\label{fig:figure1}
 \end{figure}

\begin{figure}
\centering
\subfloat[\label{fig:a}]{
\includegraphics[width=3.5cm]{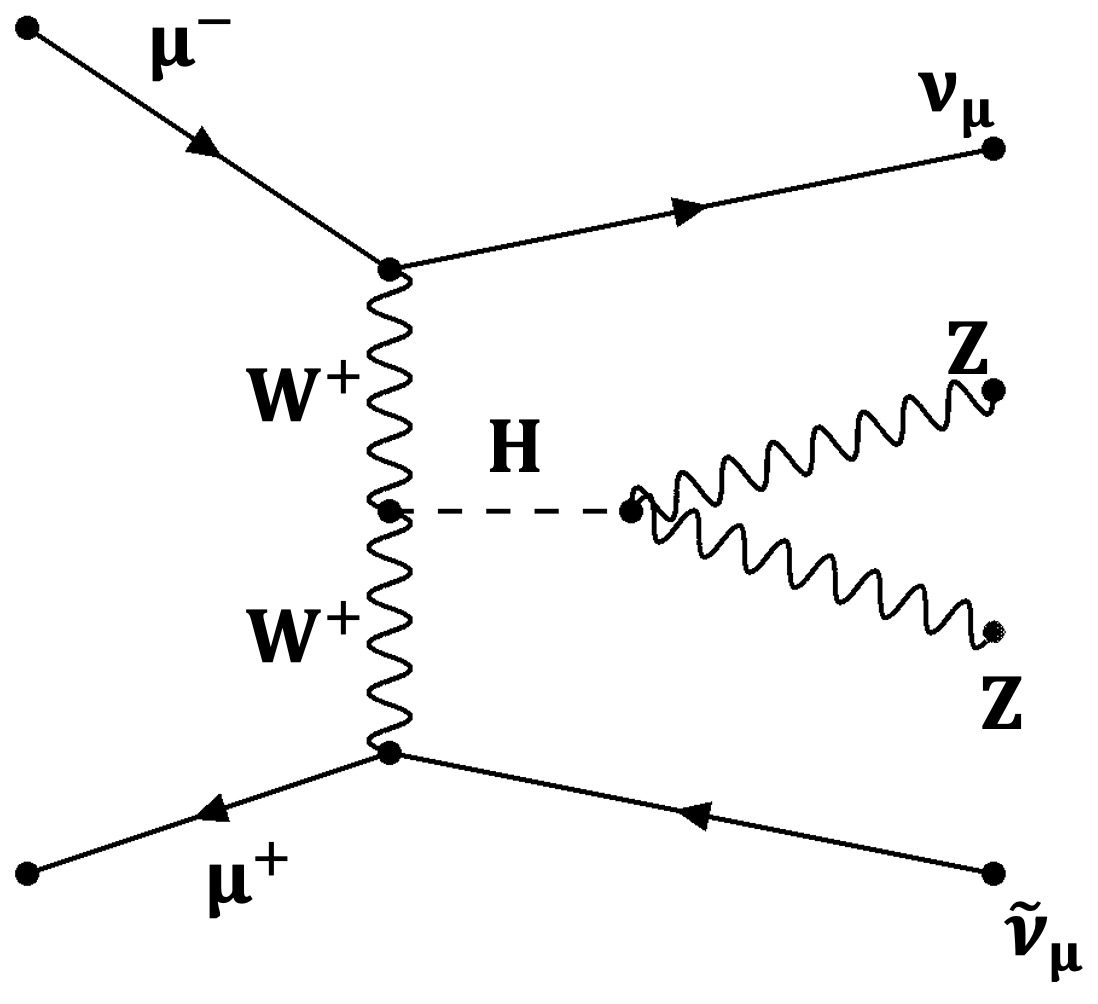}}
\subfloat[\label{fig:b}]{
\includegraphics[width=3.1cm]{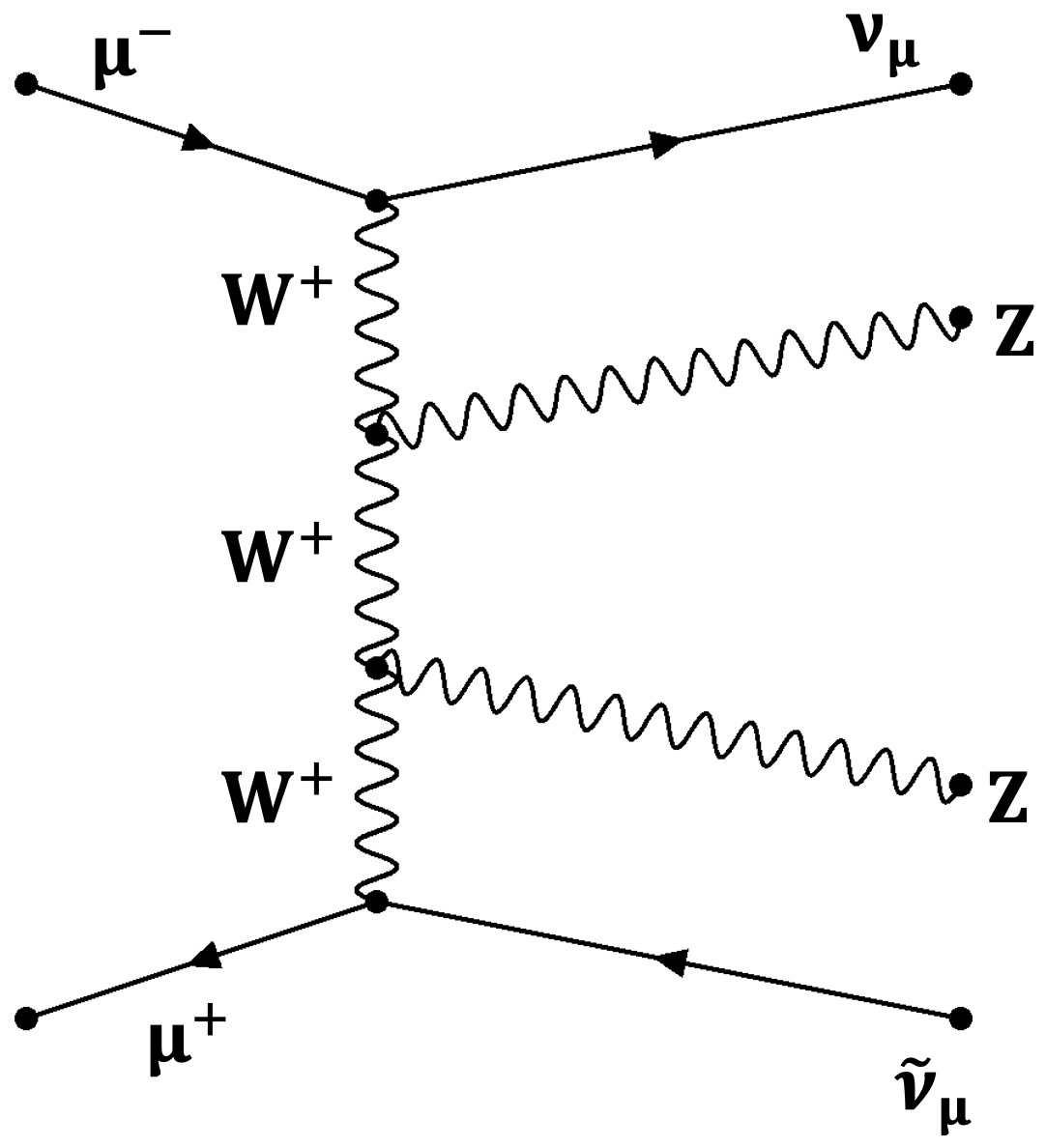}}
\\
\subfloat[\label{fig:c}]{
\includegraphics[width=3.5cm]{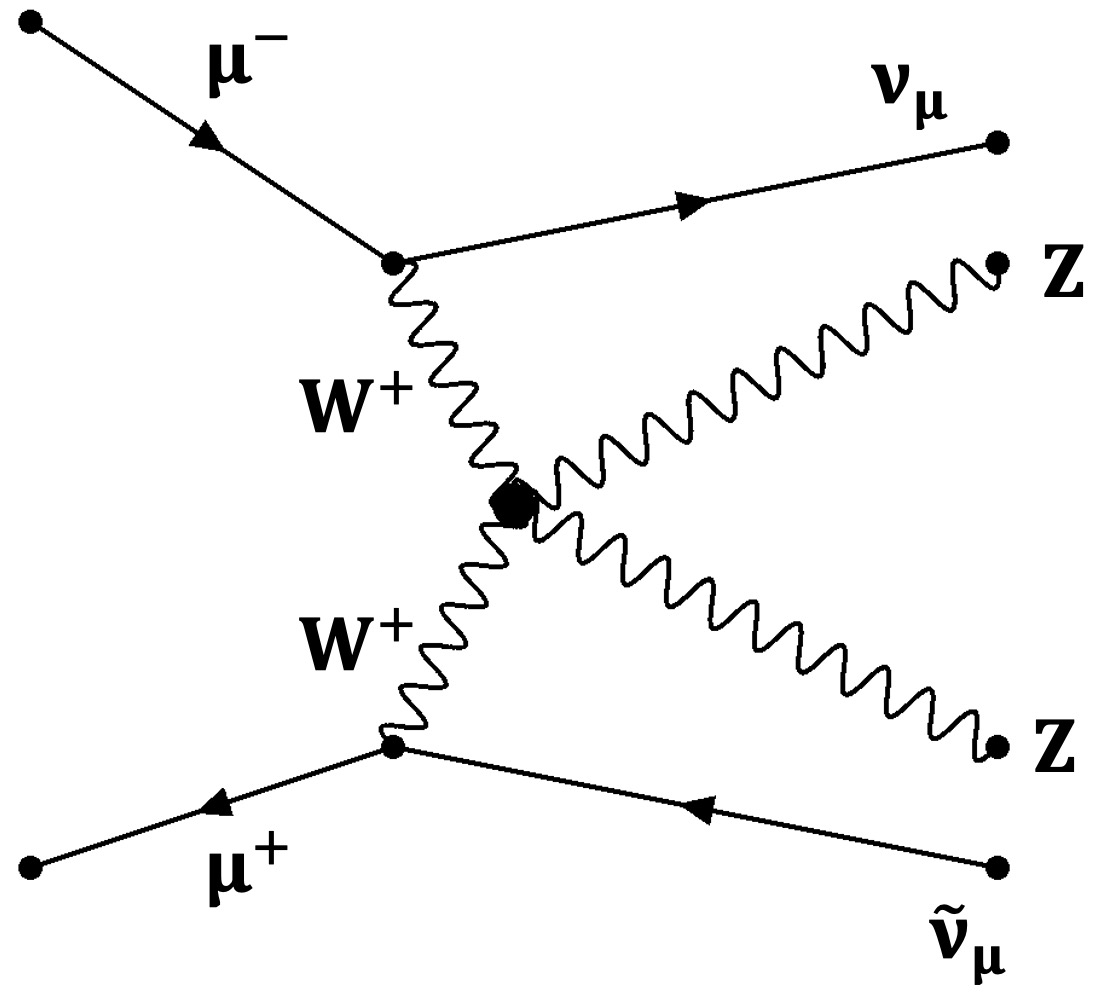}}
\caption{Example Feynman diagrams of VBS $\PZ\PZ$ production processes at a muon collider: (a) and (b) are from the SM, and (c) involves quartic gauge couplings. }
\label{fig:figure2}
 \end{figure}

In this study, we are interested in multi-Z productions which are rare processes yet to be observed. On the other hand, BSM may introduce measurable contributions and result in deviations from the SM prediction. Example processes related to $\PZ\PZ\PZ$ production in the SM and from the aQGC operator are listed in Fig.~\ref{fig:figure1}, while those for VBS $\PZ\PZ$ production are shown in Fig.~\ref{fig:figure2}
 
\section{\label{sec:level3}Simulation and Analysis Framework}

 Both the signal and background events are generated with \MGMCatNLO~\cite{Frederix:2012ps,Alwall:2014hca} at the parton-level, then showered and hadronized through \textsc{Pythia} 8.3 ~\cite{Bierlich:2022pfr}. The effects of aQGC operators are simulated with \MGMCatNLO~ using the Universal FeynRules Output module~\cite{feynrules, ufo}. The SM processes are simulated with the default SM model. \delphes~\cite{deFavereau:2013fsa} version 3.0 is used to simulate detector effects with the settings for the muon collider detector~\cite{mucard}. Jets are clustered from the reconstructed stable particles (except electrons and muons) using \textsc{FastJet}~\cite{Cacciari:2011ma} with the $k_{T}$ algorithm with a fixed cone size of $R_{jet} = 0.5$. 

\begin{figure}[ht!]
    \centering
    \includegraphics[width=7cm,height=4cm]{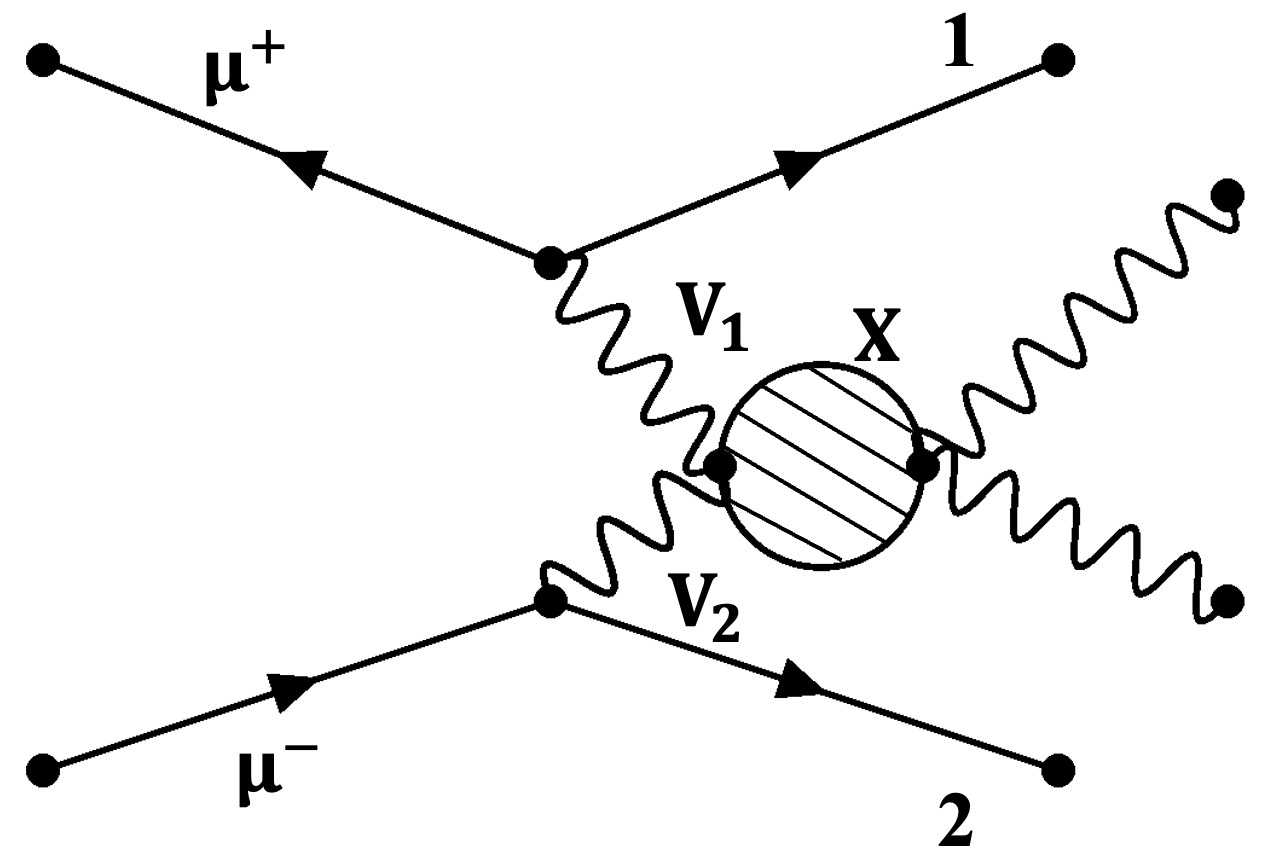}
    \caption{Example diagram of VBS processes at the muon collider. }
    \label{fig:my_figure2}
\end{figure}

Two collider scenarios and benchmarks for multi-Z productions are considered: 1) a COM energy of $\sqrt{s} = 1\,\PTeV$ for $\PZ\PZ\PZ$ direct productions, and 2) a $10\,\PTeV$ scale muon collider, where VBS~\cite{Costantini:2020stv} is the dominate production mechanism, with a more general VBS Feynman diagram as shown in Fig.~\ref{fig:my_figure2}, which includes our VBS $\PZ\PZ$ signal process. Both scenarios are studied under an integral luminosity of $10\,\abinv$.

In the study of the tri-Z boson production at a muon collider, we focus on either a pure leptonic decay: $\mu^{+}\mu^{-} \rightarrow\PZ\PZ\PZ \to \ell_{1}^{+}\ell_{1}^{-}\ell_{2}^{+}\ell_{2}^{-} \nu_{3}\bar{\nu}_{3}$, or a semi-leptonic decay: $\mu^{+}\mu^{-} \rightarrow\PZ\PZ\PZ \to \ell_{1}^{+}\ell_{1}^{-} \ell_{2}^{+}\ell_{2}^{-} jj$, where $\ell$ denotes electron or muon and $j$ denotes jet. In the study of the $\PZ\PZ$ productions through VBS, we consider pure-leptonic channel: $\mu^{+}\mu^{-} \rightarrow \PZ\PZ\nu_{\mu}\bar{\nu}_{\mu} \rightarrow 4\ell+\nu_{\mu}\bar{\nu}_{\mu}$ and semi-leptonic channel: $\mu^{+}\mu^{-} \rightarrow \PZ\PZ\nu_{\mu}\bar{\nu}_{\mu}\rightarrow 2\ell 2j+\nu_{\mu}\bar{\nu}_{\mu}$.  
The interference effect is included in our simulations with MadGraph.
Backgrounds are classified into several categories: 
\begin{itemize}
    \item  P1: s-channel processes: 
    \begin{itemize}
    \item $\mu^{+}\mu^{-}\rightarrow X=at\bar{t}+bV+c\PH$, with a, b, c as integers. 
    \end{itemize}
    \item  P2: VBS processes further divided into~\cite{Yang:2021zak}: 
    \begin{itemize}
    \item P2.1: $\PW^{+}\PW^{-}$ fusion with two neutrinos in the final state, denoted as $\PW\PW\_{\rm VBS}$ below. 
    \item P2.2 $ZZ/Z\gamma/\gamma\gamma$ fusion with two muons in the final state, denoted as $\PZ\PZ\_{\rm VBS}$ below. 
    \item P2.3: $\PW^{\pm}\PZ/\PW^{\pm}\gamma$ fusion with one muon and one neutrino in the final state, denoted as $\PW\PZ\_{\rm VBS}$ below. 
    \end{itemize}
\end{itemize}

We list all considered backgrounds in Table.~\ref{tab:my_table1}:

\begin{table}[h]
     \centering
     \caption{Summary of backgrounds of the $\PZ\PZ\PZ$ process in this study.}
     \label{tab:my_table1}
     \begin{tabular}{p{2.5cm}<{\centering}p{6cm}<{\centering}}
     \hline
     SM process type & selected backgrounds \\
     \hline
     P1: s-channel & $\PH t\overline{t}, \PZ t\overline{t}, \PW\PW t\overline{t},\PZ\PZ\PH$, $\PZ\PH\PH, \PW\PW\PZ, \PH\PH,$ \\ & $ \PW\PW\PH, \PW\PW\PW\PW, \PW\PW\PZ\PH, \PW\PW\PZ\PZ$ \\
     P2.1: $\PW\PW\_{\rm VBS}$ & $t\overline{t}, \PW\PW\PH, \PZ\PH\PH, \PZ\PZ\PH$, $\PZ\PZ\PZ, \PW\PW\PZ, \PH\PH,$ \\ & $\PZ\PZ, \PZ\PH$ \\
     P2.2: $\PZ\PZ\_{\rm VBS}$ & $\PW\PW, \PZ\PH, \PZ\PZ, t\overline{t}, \PZ, \PH, \PW\PW\PZ$ \\
     P2.3: $\PW\PZ\_{\rm VBS}$ & $\PW\PZ, \PW\PZ\PH, \PW\PH, \PW\PW\PW, \PW\PZ\PZ$  \\
     \hline
 \end{tabular}
 \end{table}

Multi-Z signals studied here suffer from a very low cross section while the backgrounds are comparatively overwhelming. It is thus necessary to apply selections to optimize signal yields while suppress backgrounds to a large extent. In this numerical analysis, we implement the cut-based method.
Some loose pre-selections are firstly applied to suppress events of no interest, then scans on each discriminating variable are performed to maximize the signal sensitivity.
In this numerical analysis, the cut-based selections are respectively optimized for each signal process.

\section{\label{sec:3Z} $\PZ\PZ\PZ$ direct productions at a $1\,\PTeV$ muon collider}
\subsection{\label{subsec:level4a}The pure-leptonic channel}
To suppress events of no interest firstly, several pre-selections are applied: the event must include exactly four leptons with transverse momentum $p_\mathrm{T, \ell}>20\PGeV$,  absolute pseudo-rapidity $|\eta_{\ell}|<2.5$, and lepton pair geometrical separation of $\Delta R_{\ell \ell} \ >0.4$ is required, where $\Delta R=\sqrt{(\Delta\phi)^{2}+(\Delta\eta)^{2}}$, the $\Delta\phi$ and the $\Delta\eta$ are the azimuthal angle separation and the pseudorapidity separation of two particles. The four leptons are classified and clustered into two reconstructed bosons $(\PZ_{1},\PZ_{2})$, with their mass denoted as $M_{\ell\ell,1}, M_{\ell\ell,2}$, following the clustering algorithm as show below:

\begin{itemize}
\item Construct all possible opposite sign lepton pairs candidates: ($\ell_{1}\ell_{2}$, $\ell_{3}\ell_{4}$), and ($\ell_{1}\ell_{4},\ell_{2}\ell_{3}$),
\item Calculate the corresponding mass difference:
\begin{equation}
\Delta{M}_{4 \ell}=|M_{\ell\ell, 1}-M_{Z}|+ | M_{\ell\ell, 2}-M_{Z}|,
\end{equation} 
\item Choose the minimum $\Delta{M}_{4 \ell}$ as the targeted lepton pairs, and we define $M_{\ell\ell, 1} > M_{\ell\ell, 2}$. 
\end{itemize}

The selections for the further optimized signal over backgrounds are listed in Table~\ref{tab:table_ZZZ_pure}, where the variable 
$M_{4\ell}$ denotes the invariant mass of the four charged leptons decaying from two Z bosons; 
$M_{\ell\ell, 1}$ and $M_{\ell\ell, 2}$ are the invariant masses of two leptons decayed from the reconstructed $\PZ_{1}$ and $\PZ_{2}$; 
$p_{T,4\ell}$ is the transverse momentum of the four leptons decaying from two Z bosons; 
$p_{T, \ell\ell, 1}$ and $p_{T, \ell\ell, 2}$ are the transverse momentum of two leptons decayed from the reconstructed $\PZ_{1}$ and $\PZ_{2}$; 
$\Delta R_{\ell\ell, 1}$ and $\Delta R_{\ell\ell, 2}$ are the geometrical separations of two leptons decayed from the reconstructed $\PZ_{1}$ and $\PZ_{2}$; 
$|\eta_{\ell\ell, 1}|$ and $|\eta_{\ell\ell, 2}|$ are the absolute pseudorapidities of the reconstructed $\PZ_{1}$ and $\PZ_{2}$;
$p_{T,\ell}^{leading}$ denotes the highest $p_{T}$ in the four charged leptons' transverse momentum; 
\met is the missing transverse energy; $M_{recoil}$ is the recoil mass of four leptons, which can be calculated as below,
\begin{equation}
    M_{recoil}=\sqrt{(\sqrt{s}-E)^{2}-P^{2}}.
\end{equation}
where $\sqrt{s}$ is the COM energy, $E$ and $P$ are the sum of detectable daughter particles' energy and the sum of detectable daughter particles' momentum. In the selections, the SM signal and the aQGC signal are optimized separately. For the SM signal, the efficiencies of pure-leptonic channel and semi-leptonic channel are 0.75 and 0.34; for the aQGC signal, the efficiencies of pure-leptonic channel and semi-leptonic channel are 0.82 and 0.40.

\begin{table}[h]
     \centering
     \caption{Event selections for the $\PZ\PZ\PZ$ in the pure-leptonic channel.}
     \label{tab:table_ZZZ_pure}
     \begin{tabular}{p{2cm}<{\centering}p{3cm}<{\centering}p{3cm}<{\centering}}
         \hline
            variables & limits for SM & limits for aQGC \\
         \hline
         $M_{4\ell}$ & $[200\PGeV,900\PGeV]$ &  $[150\PGeV,910\PGeV]$  \\
         $M_{\ell\ell, 1}$ & $[80\PGeV,120\PGeV]$  &  $[70\PGeV,130\PGeV]$  \\
         $M_{\ell\ell, 2}$ & $[60\PGeV,100\PGeV]$  & $[40\PGeV,100\PGeV]$ \\
         $p_{T,4\ell}$ & $[30\PGeV,480\PGeV]$  &  $[30\PGeV,500\PGeV]$  \\
         $p_{T, \ell\ell, 1}$ & $<500\PGeV$  & $<500\PGeV$ \\
         $p_{T, \ell\ell, 2}$ & $<460\PGeV$  & $<500\PGeV$ \\ 
         $\Delta R_{\ell\ell, 1}$ & $[0.4,3.3]$  & $[0.4,3.1]$ \\
         $\Delta R_{\ell\ell, 2}$ & $[0.4,3.3]$  & $[0.4,3.1]$ \\
         $|\eta_{\ell\ell, 1}|$ & $<2.5$  & $<2.5$ \\
         $|\eta_{\ell\ell, 2}|$ & $<2.5$  & $<2.5$ \\
         $p_{T, \ell}^{leading}$ & $[20\PGeV,380\PGeV]$  & $[25\PGeV,460\PGeV]$ \\
         $\Delta{M}_{4 \ell}$ & $<20\PGeV$  &  $<50\PGeV$  \\
         $\met$ & $[50\PGeV,460\PGeV]$  & $[100\PGeV,480\PGeV]$ \\
         $M_{recoil}$ & $<300\PGeV$  & $[35\PGeV,225\PGeV]$ \\
        \hline
     \end{tabular}
 \end{table}

\begin{figure}
\centering
\subfloat[\label{fig:a}]{
\includegraphics[width=4.5cm,height=4cm]{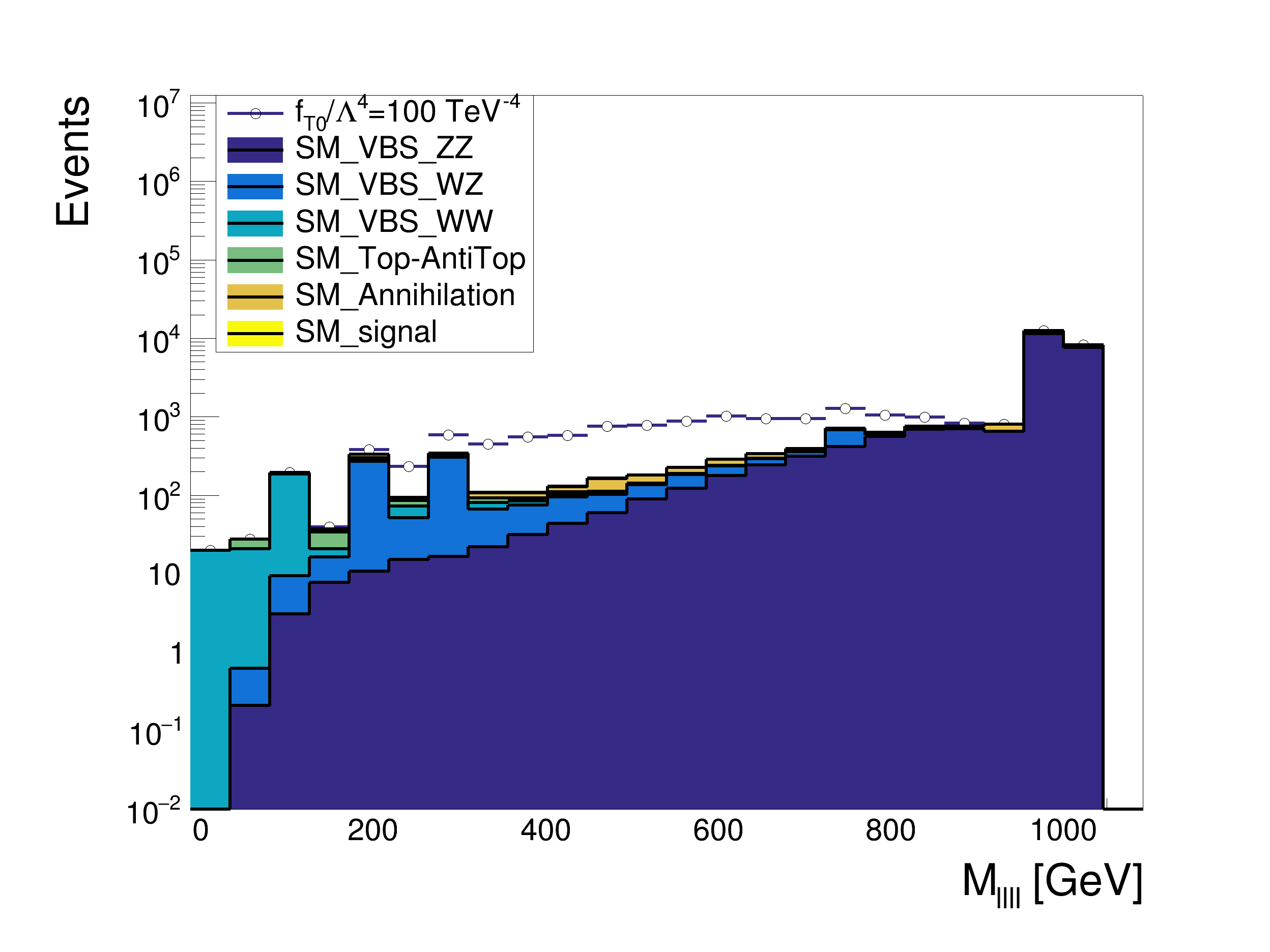}}
\subfloat[\label{fig:b}]{
\includegraphics[width=4.5cm,height=4cm]{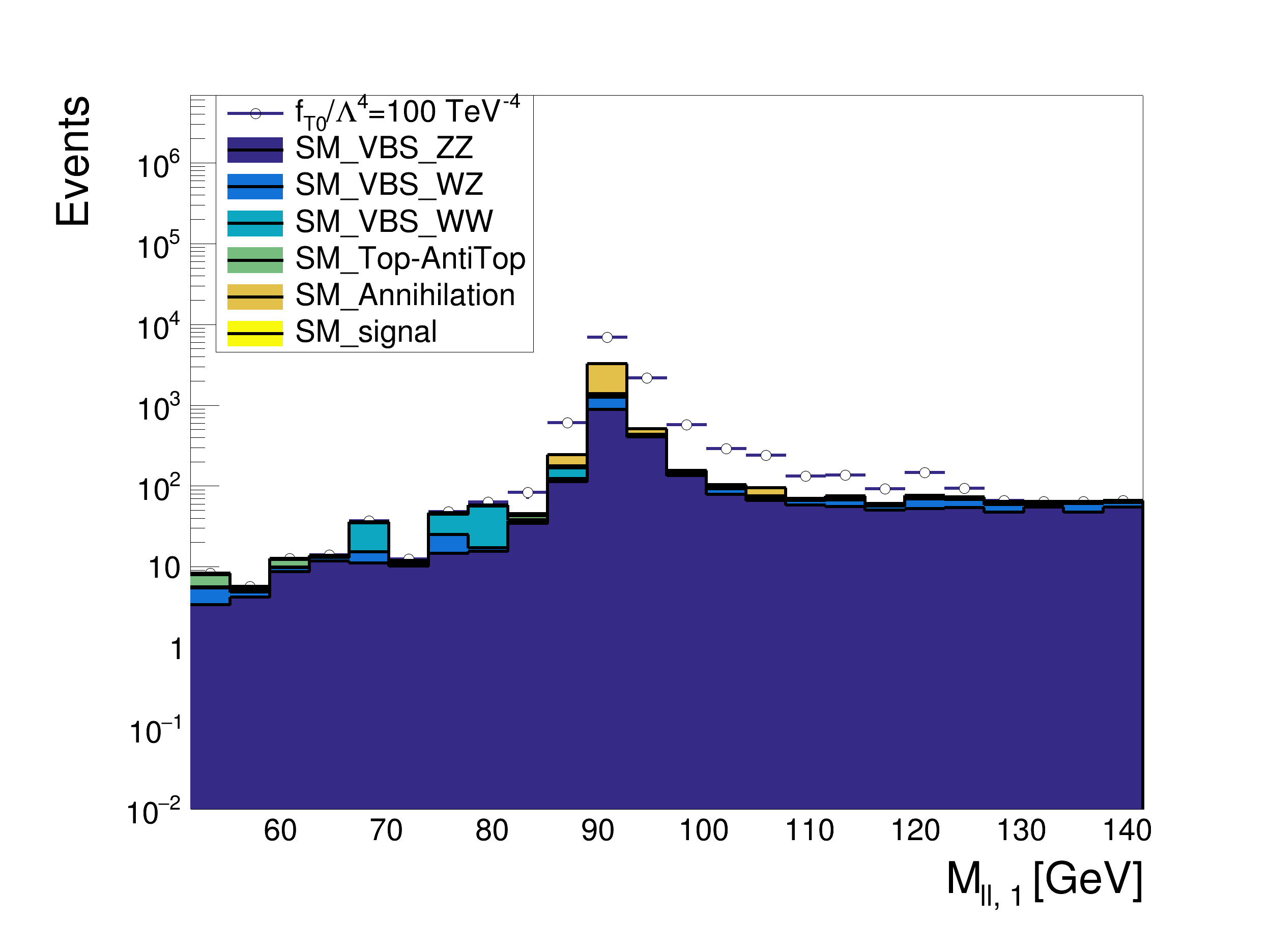}}
\\
\subfloat[\label{fig:c}]{
\includegraphics[width=4.5cm,height=4cm]{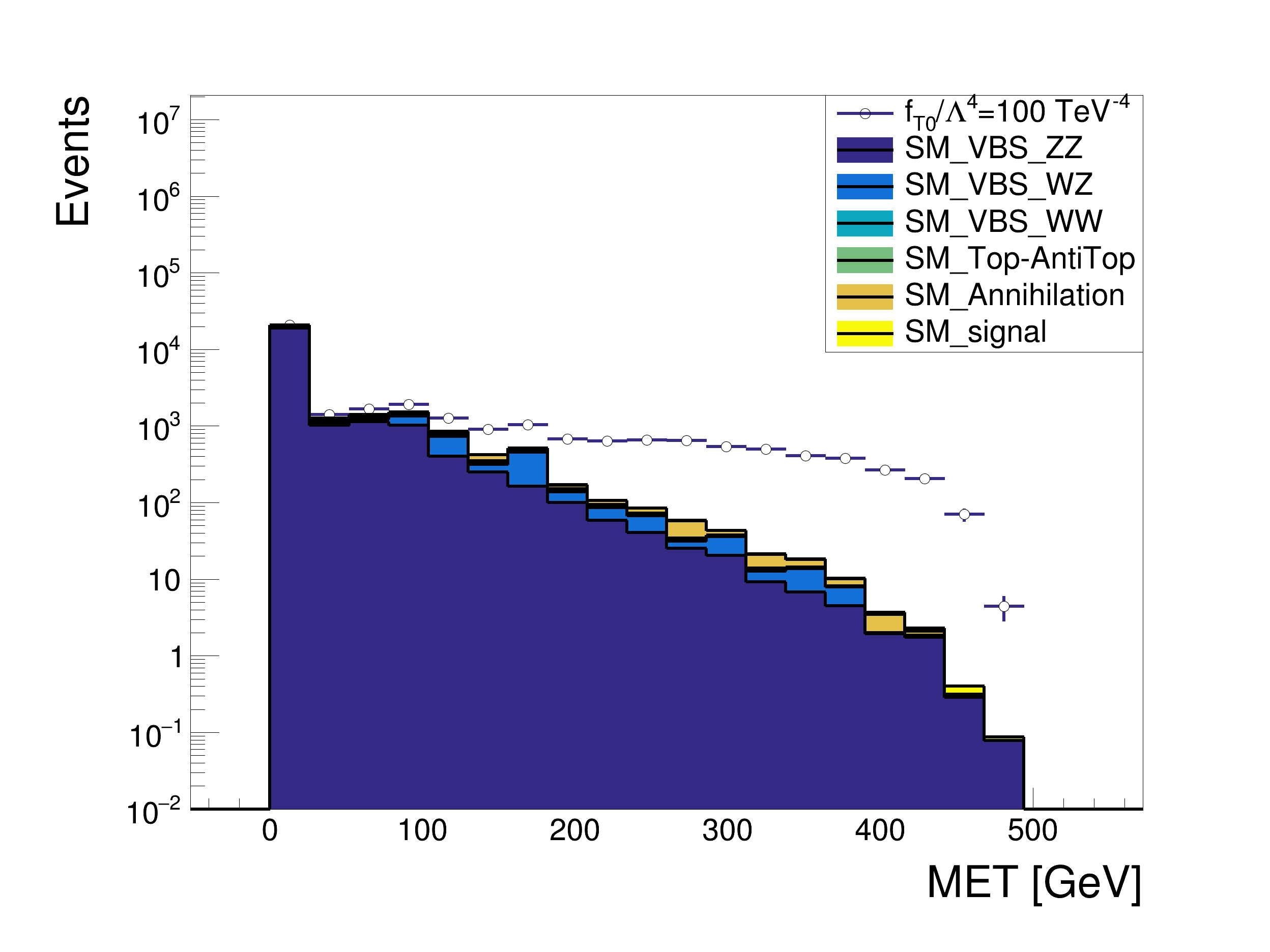}}
\subfloat[\label{fig:d}]{
\includegraphics[width=4.5cm,height=4cm]{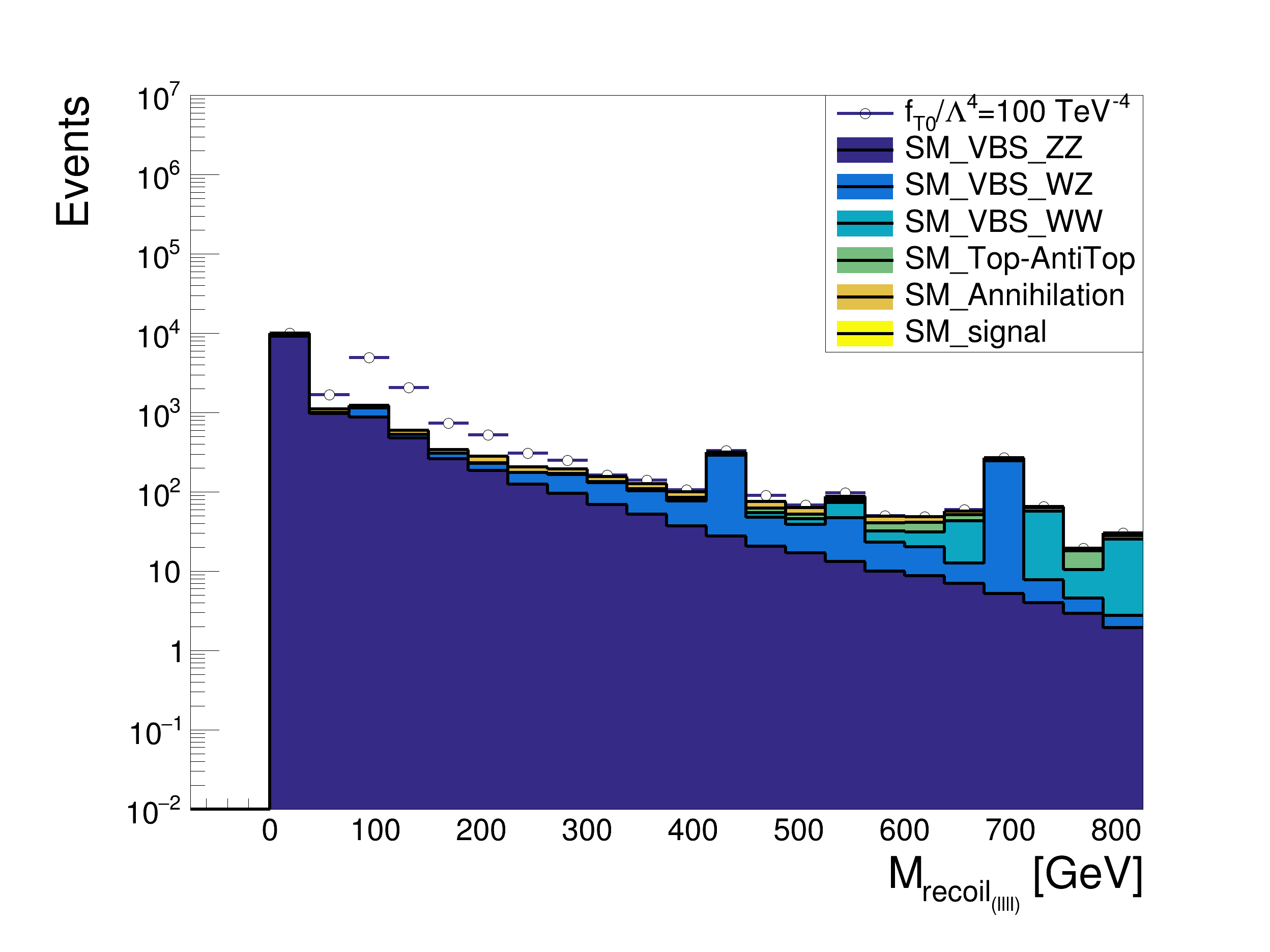}}
\caption{Various distributions for the $\PZ\PZ\PZ$ direct productions in the pure-leptonic channel, at a muon collider of $\sqrt{s}=1\,\PTeV$  and $\mathcal{L}=10\,\abinv$. (a) invariant mass of four leptons, $M_{4\ell}$, (b)invariant mass of two leptons, $M_{\ell\ell, 1}$, (c) missing transverse energy, $\met$, and (d) the recoil mass of four leptons, $M_{recoil}$.}
\label{fig:figure3}
 \end{figure}

Fig.~\ref{fig:figure3} shows some typical distributions before all selections, including $M_{4\ell}$, $M_{\ell\ell,1}$, \met, and $M_{recoil}$.
We find that both $M_{\ell\ell, i} (i = 1, 2)$ and $M_{recoil}$ can distinguish signal and backgrounds well. For the SM signal, we obtain the significance~\cite{Cowan:2010js}: $\sqrt{2((s+b)\ln (1+s/b)-s)} = 0.9\sigma$, with $S$ and $B$ as signal and background yields, respectively. 
The yield of the process is calculated by summing up the selected events' weights, which are obtained through: $\sigma\times\mathcal{L}\over{N}$, $\sigma$ is the cross-section ot the sample, $\mathcal{L}$ is the luminosity, and N is the total of generated events.   
In these plots, we also add curves for non-zero aQGCs, taking $f_{T, 0}/\Lambda^{4} = 100\,\PTeV^{-4}$as a benchmark. The aQGCs in general lead to excess at high energy tails.

\subsection{\label{subsec:level4b}Semi-leptonic channels}

\begin{figure}
\centering
\subfloat[\label{fig:a}]{
\includegraphics[width=4.5cm,height=4cm]{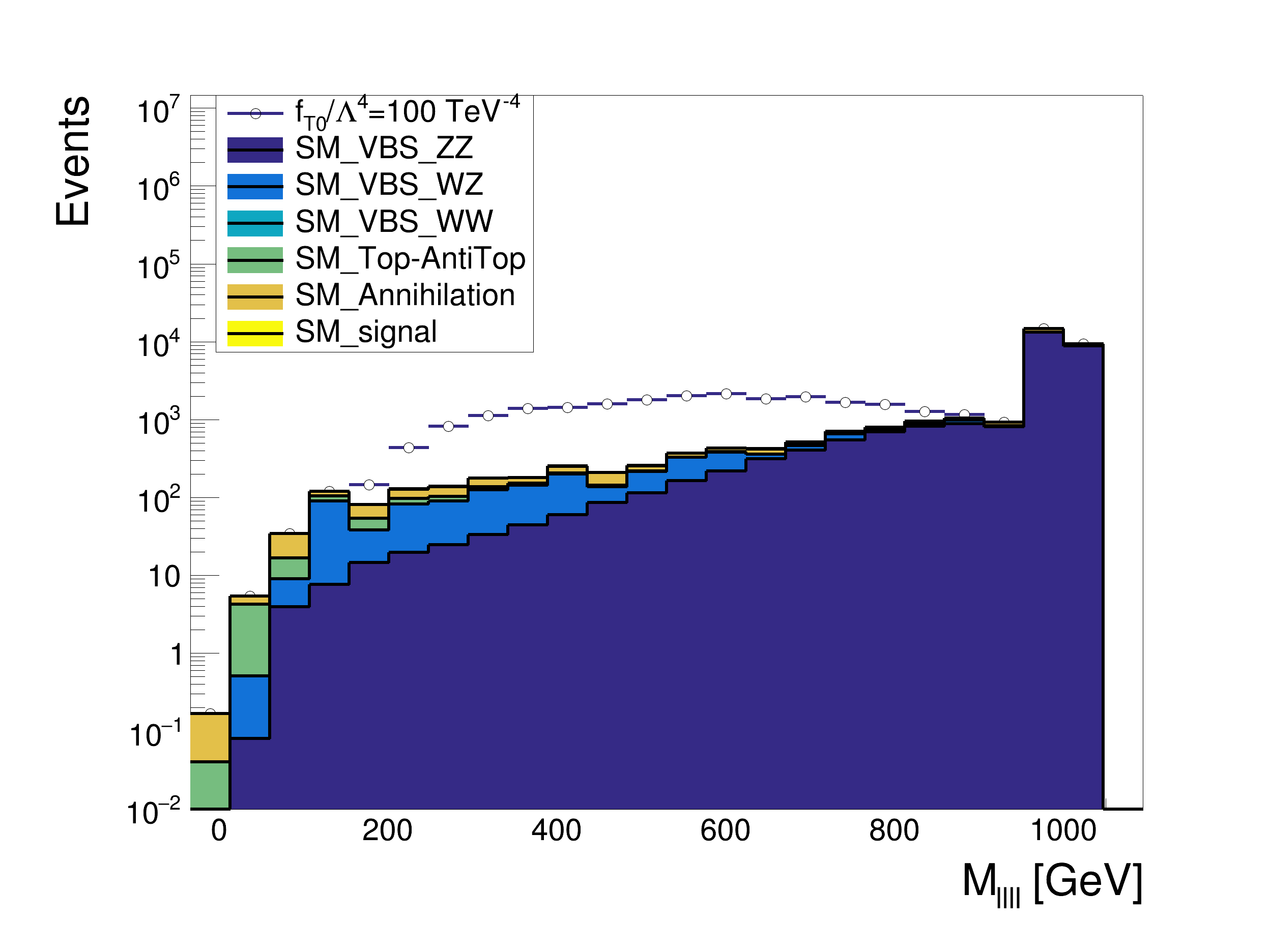}}
\subfloat[\label{fig:b}]{
\includegraphics[width=4.5cm,height=4cm]{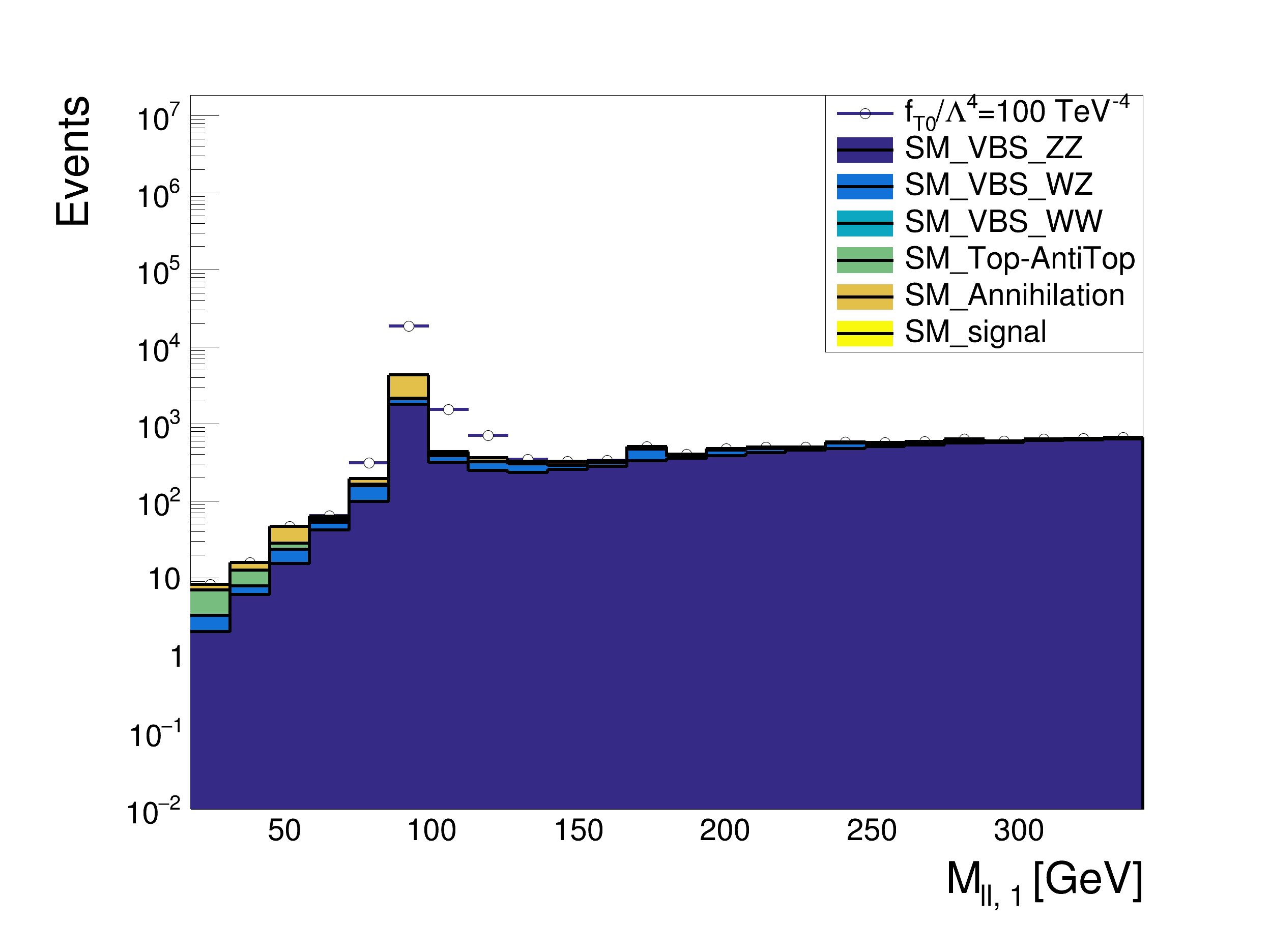}}
\\
\subfloat[\label{fig:c}]{
\includegraphics[width=4.5cm,height=4cm]{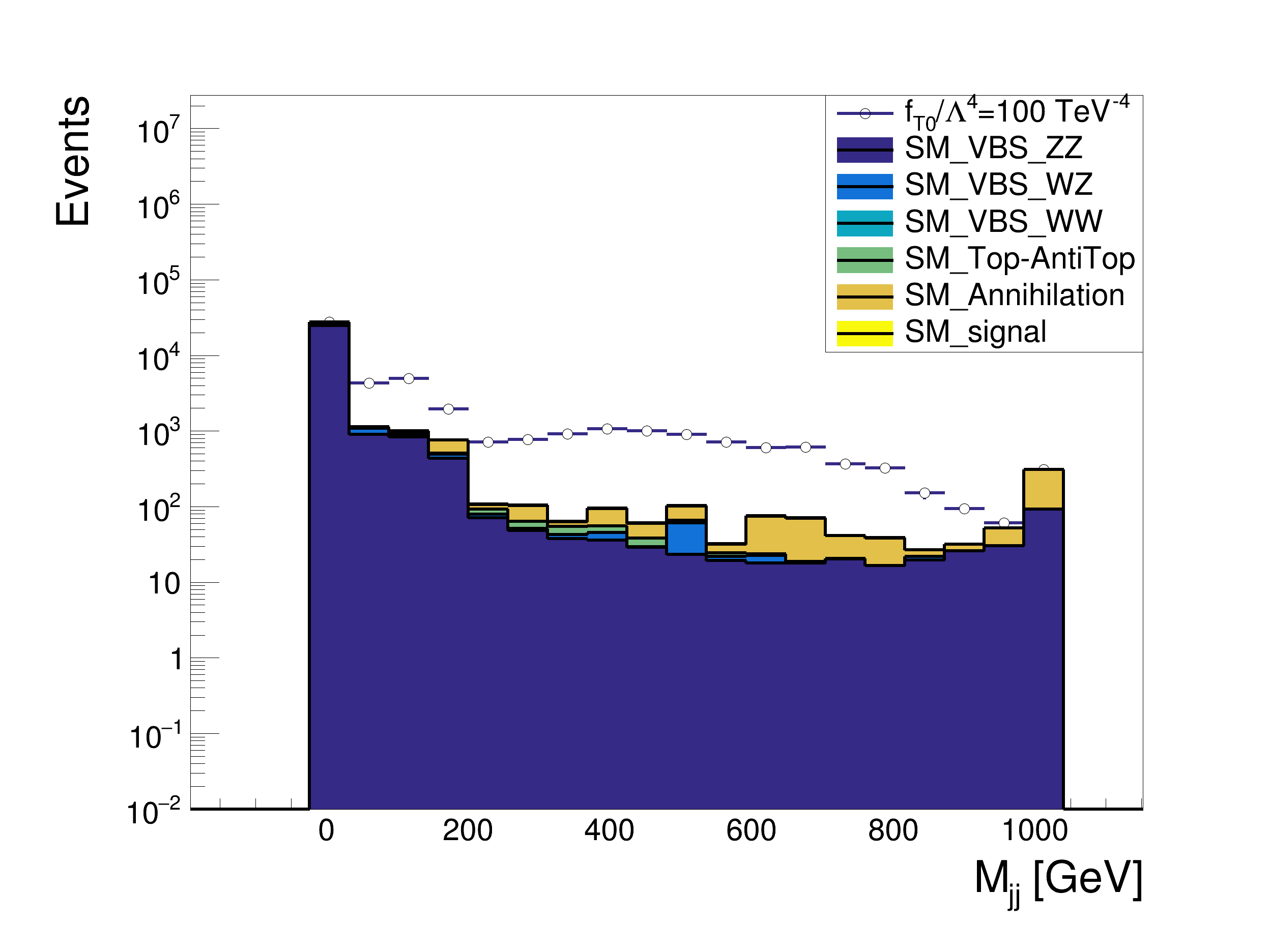}}
\subfloat[\label{fig:d}]{
\includegraphics[width=4.5cm,height=4cm]{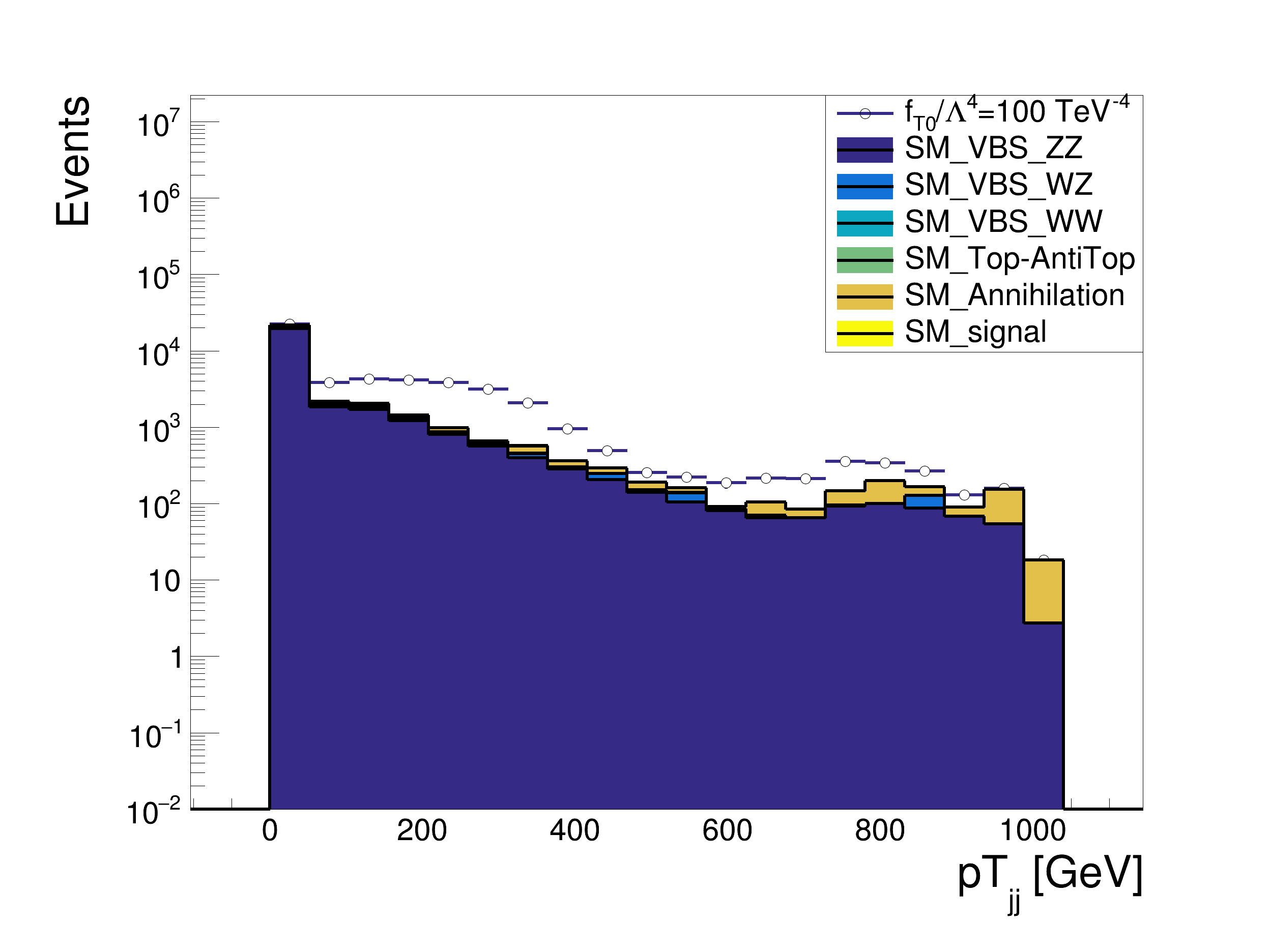}}
\caption{Various distributions for the $\PZ\PZ\PZ$ direct productions in the semi-leptonic channel, at a muon collider of $\sqrt{s}=1\,\PTeV$ and $\mathcal{L}=10\,\abinv$. (a) invariant mass of four leptons $M_{4\ell}$ distribution, (b)invariant mass of two leptons $M_{\ell\ell, 1}$, (c) invariant mass of two jets $M_{jj}$, and (d) the transverse momentum of two jets in final state $p_{T,jj}$ distribution.}
\label{fig:figure4}
 \end{figure}

Similar analysis is applied for the semi-leptonic channel, $\PZ\PZ\PZ \rightarrow 4\ell+2jets$. Fig.~\ref{fig:figure4} shows distributions of $M_{4\ell}$, $M_{\ell\ell, 1}$, the invariant mass of two jets decayed from the other Z boson $M_{jj}$, and the transverse momentum of jet pair, $p_{T,jj}$.
 
The selections of semi-leptonic channel are listed in Table ~\ref{tab:table3}, 
where $\Delta R_{jj}$ is the geometrical separation of two jets; 
$|\eta_{jj}|$ is the absolute pseudorapidity of the Z boson reconstructed from two jets;
$p_{T, j}^{leading}$ denotes the highest $p_{T}$ in the two jets' transverse momentum.
 \begin{table}[h]
     \centering
     \caption{Event selections for the $\PZ\PZ\PZ$ in the semi-leptonic channel.}
     \label{tab:table3}
     \begin{tabular}{p{2cm}<{\centering}p{3cm}<{\centering}p{3cm}<{\centering}}
         \hline
            variables & limits for SM & limits for aQGC \\
         \hline
         $M_{4\ell}$ & $[200\PGeV,840\PGeV]$   & $[150\PGeV,930\PGeV]$ \\
         $M_{\ell\ell, 1}$ & $[80\PGeV,120\PGeV]$  & $[85\PGeV,130\PGeV]$ \\
         $M_{\ell\ell, 2}$ & $[60\PGeV,100\PGeV]$  & $[65\PGeV,115\PGeV]$ \\
         $M_{jj}$ & $<150\PGeV$  & $[30\PGeV,150\PGeV]$ \\
         $p_{T,4\ell}$ & $[30\PGeV,450\PGeV]$ & $[30\PGeV,480\PGeV]$ \\
         $p_{T, \ell\ell, 1}$ & $<500\PGeV$  & $<480\PGeV$ \\
         $p_{T, \ell\ell, 2}$ & $<460\PGeV$  & $<480\PGeV$ \\ 
         $p_{T, jj}$ & $<420\PGeV$  & $<500\PGeV$ \\ 
         $\Delta R_{\ell\ell, 1}$ & $[0.4,3.1]$  & $[0.4,3.3]$ \\
         $\Delta R_{\ell\ell, 2}$ & $[0.4,3.1]$  & $[0.4,3.3]$ \\
         $\Delta R_{jj}$ & $[0.4,4.0]$  & $[0.4,3.5]$ \\
         $|\eta_{\ell\ell, 1}|$ & $<2.5$  & $<2.5$ \\
         $|\eta_{\ell\ell, 2}|$ & $<2.5$  & $<2.5$ \\
         $|\eta_{jj}|$ & $<5.0$  & $<5.0$ \\
         $p_{T, \ell}^{leading}$ & $[20\PGeV,400\PGeV]$  & $[20\PGeV,420\PGeV]$ \\
         $p_{T, j}^{leading}$ & $[30\PGeV,480\PGeV]$  & $[30\PGeV,510\PGeV]$ \\
         $\Delta{M}_{4 \ell}$ & $<20\PGeV$  & $<30\PGeV$  \\
         $\met$ & $<100\PGeV$  & $<150\PGeV$ \\
         $M_{recoil}$ & $<300\PGeV$  & $[35\PGeV,225\PGeV]$ \\
        \hline
     \end{tabular}
 \end{table}

The selections improve the significance of both SM and aQGC signals. With $10\,\abinv$ of integrated luminosity at $\sqrt{s}=1\,\PTeV$, the expected yields of SM signal and background after the selections are listed in Table~\ref{tab:ZZZ_yields_SM}; 
and with the aQGC benchmark $f_{T, 0}/\Lambda^{4} = 100\,\PTeV^{-4}$, the expected yields of aQGC signal and background after the selections are listed in Table~\ref{tab:ZZZ_yields_aQGC}.
\begin{table}[h]
     \centering
     \caption{The expected yields of SM signal and background after the selections, in the $\PZ\PZ\PZ$ direct productions.}
     \label{tab:ZZZ_yields_SM}
     \begin{tabular}{p{2.5cm}<{\centering}p{2cm}<{\centering}p{3cm}<{\centering}}
         \hline
         Channels ($\sqrt{s}=1\,\PTeV$) & Expected signal yield [events]& Expected background yield [events]\\
         \hline
         Pure-leptonic chanel          & 5.18      & 31.72 \\
         Semi-leptonic chanel          & 4.48      & 5.46 \\
        \hline
     \end{tabular}
 \end{table}
\begin{table}[h]
     \centering
     \caption{The expected yields of aQGC signal and background after the selections in the $\PZ\PZ\PZ$ direct productions.}
     \label{tab:ZZZ_yields_aQGC}
     \begin{tabular}{p{2.5cm}<{\centering}p{2cm}<{\centering}p{3cm}<{\centering}}
         \hline
         Channels ($\sqrt{s}=1\,\PTeV$) & Expected signal yield [events] & Expected background yield [events]\\
         \hline
         Pure-leptonic chanel          & 5514.90      & 56.80 \\
         Semi-leptonic chanel          & 6271.79      & 9.16 \\
        \hline
     \end{tabular}
 \end{table}

After selections, the significance for this semi-leptonic channel can reach $1.7\sigma$ for the SM signal process. We further combine%~\cite{combine} 
the pure-leptonic channel and semi-leptonic resulting a higher significance of $1.9\sigma$ for the SM signal. We also provide searches for aQGCs,  and obtain the constraint range of all coefficients $f_{S,M,T}$, which will be shown in Table~\ref{tab:table4}.

\section{\label{sec:aQGC} VBS $\PZ\PZ$ productions at a $10\,\PTeV$ muon collider}

For VBS $\PZ\PZ$ process, we perform similar simulation studies as for the $\PZ\PZ\PZ$ process. We consider pure-leptonic channel: $\mu^{+}\mu^{-} \rightarrow \PZ\PZ\nu_{\mu}\bar{\nu}_{\mu} \rightarrow 4\ell+\nu_{\mu}\bar{\nu}_{\mu}$ and semi-leptonic channel: $\mu^{+}\mu^{-} \rightarrow \PZ\PZ\nu_{\mu}\bar{\nu}_{\mu}\rightarrow 2\ell 2j+\nu_{\mu}\bar{\nu}_{\mu}$. The backgrounds are also divided into P1: s-channel, P2.1: $\PW\PW\_{\rm VBS}$, P2.2: $\PZ\PZ\_{\rm VBS}$, P2.3: $\PW\PZ\_{\rm VBS}$, They are listed in Table~\ref{tab:my_table_ZZ}:

\begin{table}[h]
     \centering
     \caption{Summary of backgrounds for the VBS $\PZ\PZ$ process.}
     \label{tab:my_table_ZZ}
     \begin{tabular}{p{2.5cm}<{\centering}p{6cm}<{\centering}}
     \hline
     SM process type & selected backgrounds \\
     \hline
     P1: s-channel & $\PW\PW, \PZ\PZ, \PZ\PH, \PH\PH, \PZ\PH\PH,$ 
     $\PZ\PZ\PZ, \PZ\PZ\PH, \PW\PW\PH,$ \\ & $\PW\PW\PZ, t\overline{t},$
     $\PH t\overline{t}, \PZ t\overline{t}, \PW\PW t\overline{t}, \PW\PW\PW\PW,$ \\ & $ \PW\PW\PZ\PH, \PW\PW\PH\PH$  
     \\
     P2.1: $\PW\PW\_{\rm VBS}$ & $\PW\PW, \PZ\PH, \PH\PH,$
     $\PW\PW\PH, \PW\PW\PZ, \PZ\PZ\PZ,$ \\ & $\PZ\PZ\PH, \PZ\PH\PH, t\overline{t}, \PZ, \PH$ 
     \\
     P2.2: $\PZ\PZ\_{\rm VBS}$ & $\PW\PW, \PZ\PH, \PZ\PZ, t\overline{t}, \PZ, \PW\PW\PH, \PW\PW\PZ, \PH, \PH\PH,$ \\ & $\PZ\PZ\PZ, \PZ\PZ\PH, \PZ\PH\PH$ 
     \\
     P2.3: $\PW\PZ\_{\rm VBS}$ & $\PW\PZ, \PW\PZ\PH, \PW\PH, \PW\PW\PW, \PW\PZ\PZ$  \\
     \hline
 \end{tabular}
 \end{table}

\subsection{\label{subsec:level4b}Pure-leptonic channel of VBS \PZ\PZ}
We apply pre-selections on the channel of $\mu^{+}\mu^{-}\rightarrow \PZ\PZ\nu_{\mu}\bar{\nu}_{\mu}\rightarrow4\ell+\nu_{\mu}\bar{\nu}_{\mu}$ at a muon collider with $\sqrt{s}=10\,\PTeV$ and $\mathcal{L}=10\,\abinv$ as the same as $\PZ\PZ\PZ$ analysis. The selections of pure-leptonic channel are listed in Table~\ref{tab:ZZ_pure}. The signal efficiency of the selections is 0.23.

\begin{table}[h]
     \centering
     \caption{Event selections for the VBS $\PZ\PZ$ in the pure-leptonic channel.}
     \label{tab:ZZ_pure}
     \begin{tabular}{p{2.5cm}<{\centering}p{4cm}<{\centering}}
         \hline
            variables & limits \\
         \hline
         $M_{4\ell}$ & $[1900\PGeV,8800\PGeV]$  \\
         $M_{\ell\ell, 1}$ & $[70\PGeV,140\PGeV]$ \\
         $M_{\ell\ell, 2}$ & $[70\PGeV,140\PGeV]$ \\
         $p_{T,4\ell}$ & $[200\PGeV,4000\PGeV]$ \\
         $p_{T,\ell\ell,1}$ & $[320\PGeV,2800\PGeV]$ \\
         $p_{T,\ell\ell,2}$ & $[280\PGeV,2600\PGeV]$ \\ 
         $\Delta R_{\ell\ell,1}$ & $[0.4,1.7]$ \\
         $\Delta R_{\ell\ell,2}$ & $[0.4,1.7]$ \\
         $|\eta_{\ell\ell, 1}|$ & $<2.5$ \\
         $|\eta_{\ell\ell, 2}|$ & $<2.5$ \\
         $p_{T,\ell}^{leading}$ & $[200\PGeV,3000\PGeV]$ \\
         $\Delta{M}_{4 \ell}$ & $<70\PGeV$ \\
         $\met$ & $[30\PGeV,4000\PGeV]$ \\
         $M_{recoli}$ & $<8000\PGeV$ \\
        \hline
     \end{tabular}
 \end{table}

\begin{figure}
\centering
\subfloat[\label{fig:a}]{
\includegraphics[width=4.5cm,height=4cm]{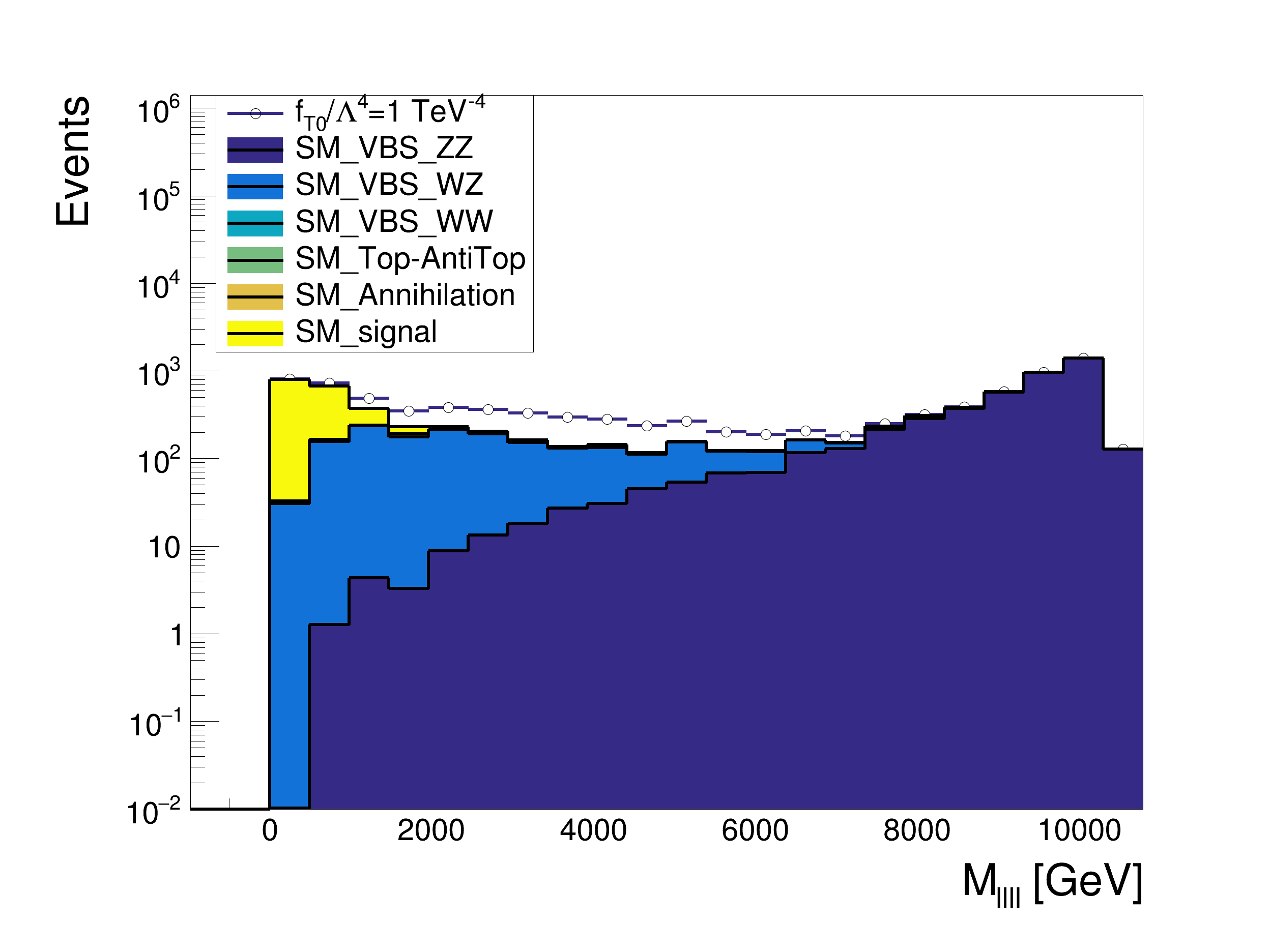}}
\subfloat[\label{fig:b}]{
\includegraphics[width=4.5cm,height=4cm]{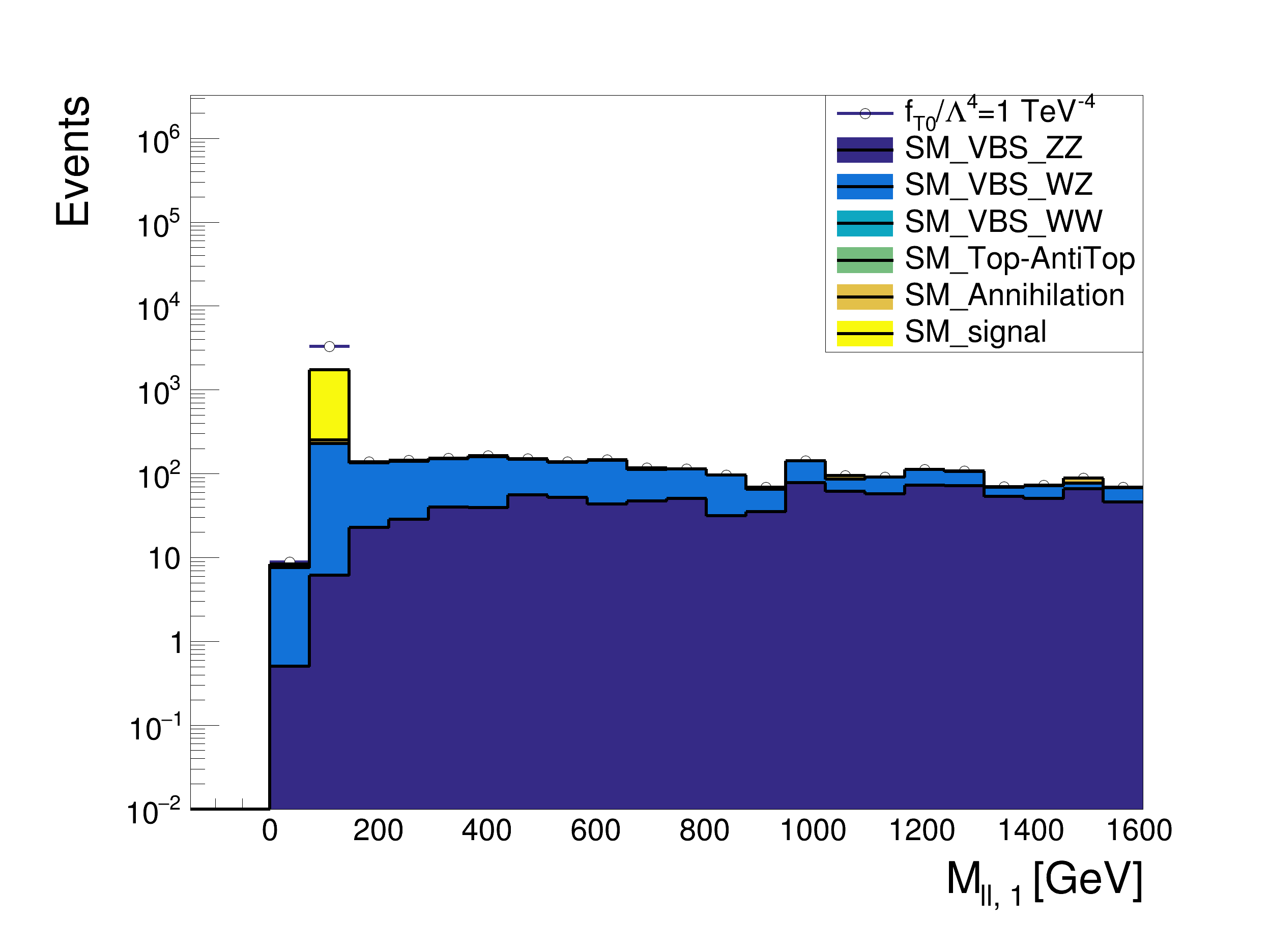}}
\\
\subfloat[\label{fig:c}]{
\includegraphics[width=4.5cm,height=4cm]{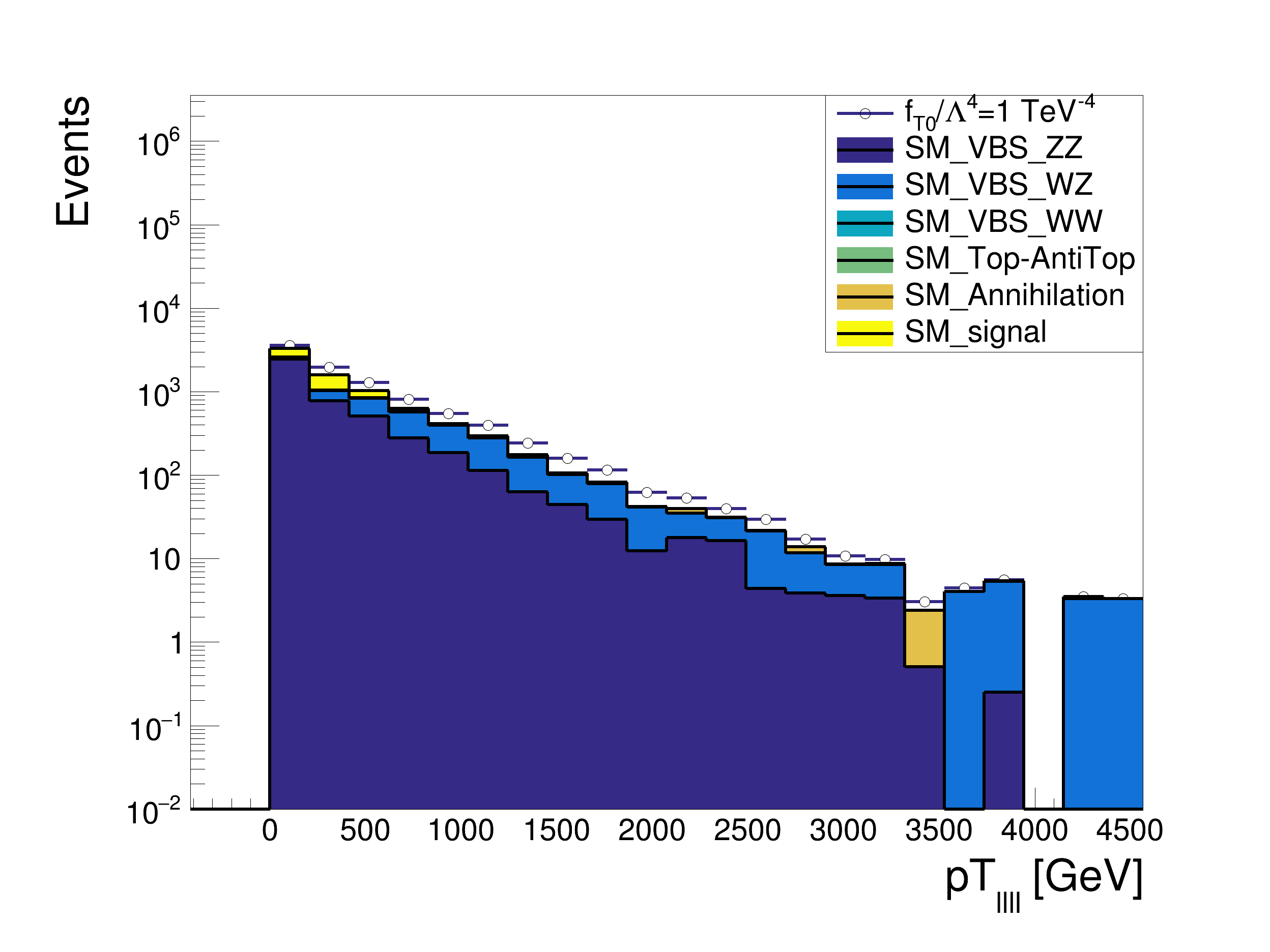}}
\subfloat[\label{fig:d}]{
\includegraphics[width=4.5cm,height=4cm]{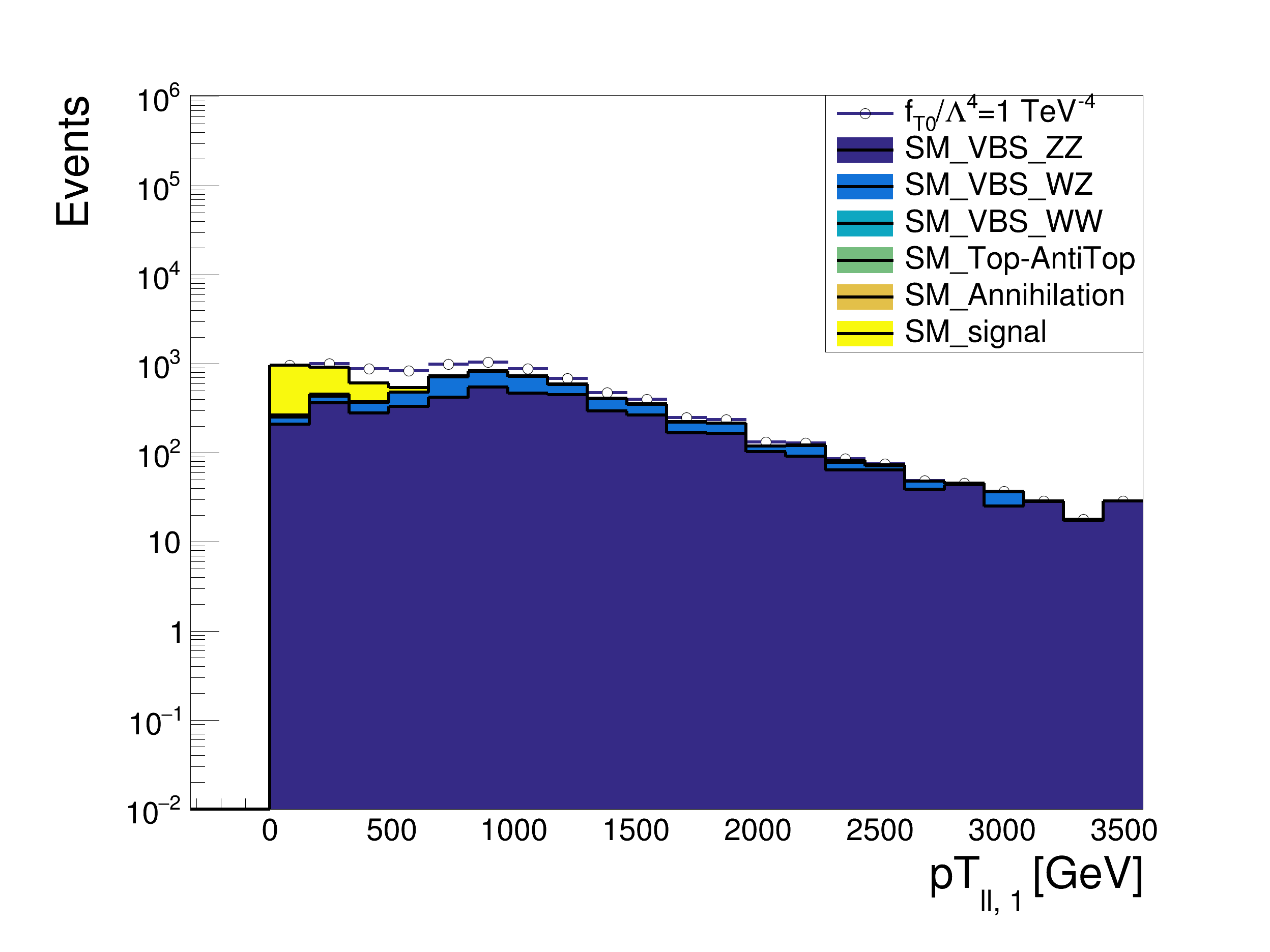}}
\caption{Simulation results of VBS $\PZ\PZ$ in the pure-leptonic channel, $\sqrt{s}=10\,\PTeV, \mathcal{L}=10\,\abinv$.
(a) invariant mass of four leptons $M_{4\ell}$ distribution, (b)invariant mass of two leptons $M_{\ell\ell, 1}$, (c) four leptons transverse momentum $p_{T,4\ell}$ distribution, and (d) two leptons' transverse momentum $p_{TT,\ell\ell,1}$ distribution.}
\label{fig:figure_ZZ}
 \end{figure}

Fig.~\ref{fig:figure_ZZ} shows the distribution of four leptons invariant mass $M_{4\ell}$, the invariant mass of two leptons $M_{\ell\ell, 1}$, the transverse momentum of four leptons in final states $p_{T,4\ell}$, and the transverse momentum of one lepton pair $p_{T,\ell\ell,1}$.  In these plots, we also add curves for non-zero aQGC, taking $f_{T, 0}/\Lambda^{4} = 1\,\PTeV^{-4}$ as a benchmark.

\subsection{\label{subsec:level4b}Semi-leptonic channels of VBS $\PZ\PZ$}
Selections in the semi-leptonic channel are listed in Table~\ref{tab:table_ZZ_semi}, 
where $M_{2\ell2j}$ is the invariant mass of two leptons and two jets decaying from two Z bosons;
$\Delta{M}_{2\ell2j}$ is the mass diference defined as:
\begin{equation}
\Delta{M}_{2\ell2j}=|M_{\ell\ell}-M_{Z}|+ | M_{jj}-M_{Z}|,
\end{equation} 
The signal efficiency of the selections is 0.033.

\begin{table}[h]
     \centering
     \caption{Event selections for the VBS $\PZ\PZ$ in the semi-leptonic channel.}
     \label{tab:table_ZZ_semi}
     \begin{tabular}{p{2.5cm}<{\centering}p{4cm}<{\centering}}
         \hline
            variables & limits \\
         \hline
         $M_{2\ell2j}$ & $[2000\PGeV,8000\PGeV]$ \\
         $M_{\ell\ell}$ & $[40\PGeV,140\PGeV]$  \\
         $M_{jj}$ & $[30\PGeV,150\PGeV]$ \\
         $p_{T,2\ell2j}$ & $[500\PGeV,8000\PGeV]$ \\
         $p_{T,\ell\ell}$ & $[200\PGeV,3000\PGeV]$ \\
         $p_{T,jj}$ & $[400\PGeV,4000\PGeV]$ \\ 
         $\Delta R_{\ell\ell}$ & $[0.4,1.7]$ \\
         $\Delta R_{jj}$ & $>0.4$ \\
         $|\eta_{\ell}|$ & $<2.5$ \\
         $|\eta_{j}|$ & $<5.0$ \\
         $p_{T,\ell}^{leading}$ & $[200\PGeV,2500\PGeV]$ \\
         $p_{T,j}^{leading}$ & $[200\PGeV,3000\PGeV]$ \\
         $\Delta{M}_{2\ell2j}$ & $<200\PGeV$ \\
         $\met$ & $[30\PGeV,3500\PGeV]$ \\
         $M_{recoli}$ & $[1000\PGeV,7000\PGeV]$ \\
        \hline
     \end{tabular}
 \end{table}
 
\begin{figure}
\centering
\subfloat[\label{fig:a}]{
\includegraphics[width=4.5cm,height=4cm]{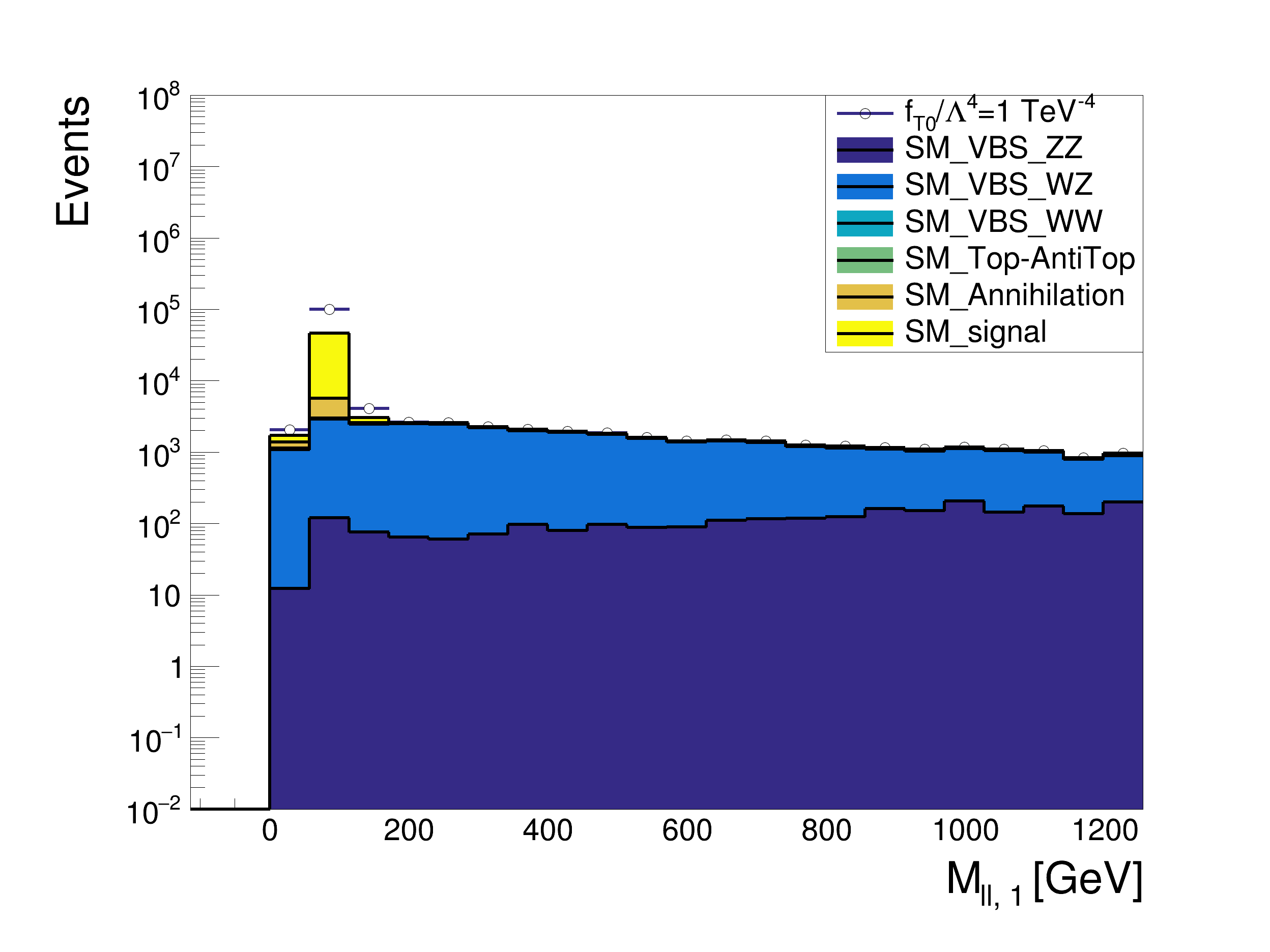}}
\subfloat[\label{fig:b}]{
\includegraphics[width=4.5cm,height=4cm]{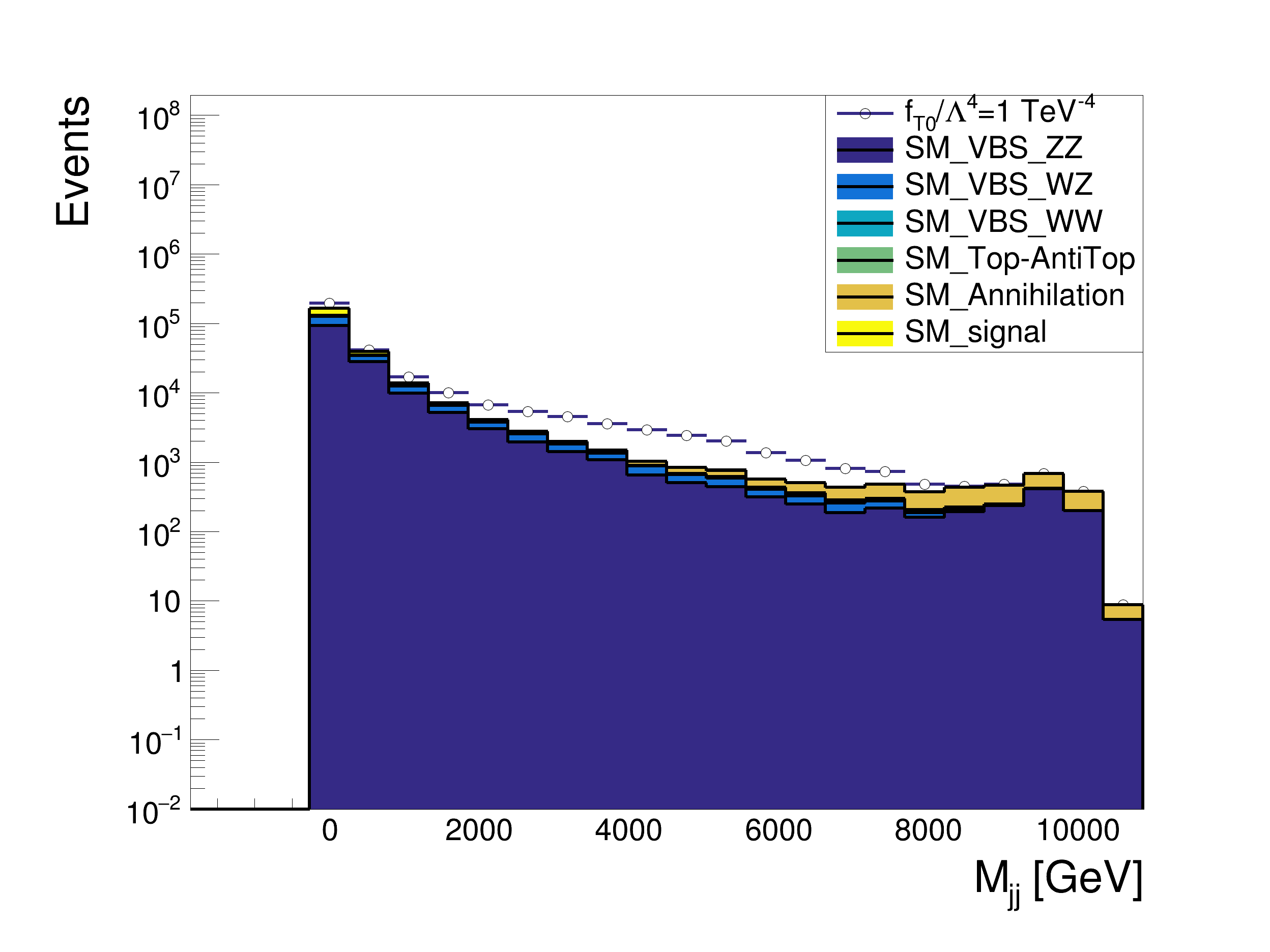}}
\\
\caption{Simulation results of VBS $\PZ\PZ$, in the semi-leptonic channel, $\sqrt{s}=10\,\PTeV, \mathcal{L}=10\,\abinv$.
(a) invariant mass of two leptons $M_{\ell\ell}$ distribution and (b) invariant mass of two leptons $M_{jj}$.}
\label{fig:figure_ZZ_jj}
 \end{figure}

Fig.~\ref{fig:figure_ZZ_jj} shows distributions of the invariant mass of jet pair $M_{jj}$, lepton pair mass $M_{\ell\ell}$, together with the aQGC signal with coefficient $f_{T,0}/\Lambda^{4}=1\,\PTeV^{-4}$.

After applying the above selections, the significance of aQGC signal is improved. With $10\,\abinv$ of integrated luminosity at $\sqrt{s}=10\,\PTeV$ and the aQGC benchmark $f_{T, 0}/\Lambda^{4} =1\,\PTeV^{-4}$, the expected yields of aQGC signal and background after the selections are listed in Table~\ref{tab:ZZ_yields}.

\begin{table}[h]
     \centering
     \caption{The expected yields of aQGC signal and background after the selections in the VBS $\PZ\PZ$ productions.}
     \label{tab:ZZ_yields}
     \begin{tabular}{p{2.5cm}<{\centering}p{2cm}<{\centering}p{3cm}<{\centering}}
         \hline
         Channels ($\sqrt{s}=10\,\PTeV$) & Expected signal yield [events] & Expected background yield [events]\\
         \hline
         Pure-leptonic chanel          & 686.97      & 0.48 \\
         Semi-leptonic chanel          & 315.40      & 0.15 \\
        \hline
     \end{tabular}
 \end{table}
 
We also obtain the limits of all aQGC coefficients of VBS $\PZ\PZ$ process, which are listed in Table~\ref{tab:table_ZZ_constrain}.

\section{\label{sec:level5}Results and Discussions}
We study multi-Z productions of $\PZ\PZ\PZ$ at a muon collider with $\sqrt{s}=1\,\PTeV$, $\mathcal{L}=10\,\abinv$. Through detailed simulation and signal background analysis, we obtain a significance for the $\PZ\PZ\PZ$ direct production in the SM as $1.9\sigma$ after combining the results from pure-leptonic and semi-leptonic channels. We also provide the constraints of aQGC coefficients~\cite{QCKM-Model} at the 95\% CL. Furthermore, high energy muon collider is an ideal place to research VBS processes, such as the VBS $\PZ\PZ$ production process. We present the distribution of various variables and summarize the constraints of aQGC coefficients. 
For $\PZ\PZ\PZ$ process, the constraints of coefficients at 95\% CL are listed in Table~\ref{tab:table4}, and for VBS $\PZ\PZ$ production process, the constraints of aQGC coefficients at 95\% CL are listed in Table~\ref{tab:table_ZZ_constrain}, the unit is $\PTeV^{-4}$. 

In ZZZ direct productions, some operators degenerate, such as: $f_{S0}$, $f_{S1}$, and $f_{S2}$; $f_{T0}$ and $f_{T1}$; $f_{T5}$ and $f_{T6}$. But we still keep all the constraints in Table~\ref{tab:table4} for the completeness of the set of operator coefficients.  

Comparing with some existing VBS $\PZ\PZ$ aQGC constraints from CMS experiment in LHC, which are based on a data sample of proton-proton collisions at COM = 13 $PTeV$ with an integrated luminosity of $\mathcal{L}=137fb^{-1}$: $f_{T, 0}/\Lambda^{4}:[-0.24,0.22],  f_{T, 1}/\Lambda^{4}:[-0.31,0.31], f_{T, 2}/\Lambda^{4}:[-0.63,0.59]$ in ~\cite{CMS:2020fqz}, our results give stronger limits: $f_{T, 0}/\Lambda^{4}:[-0.11, 0.082], f_{T, 1}/\Lambda^{4}:[-0.14,0.11], f_{T, 2}/\Lambda^{4}:[-0.27,0.21]$.  
\begin{table}[h]
     \centering
     \caption{Limits at the 95\% CL on aQGC coefficients for the $\PZ\PZ\PZ$ process.}
     \label{tab:table4}
     \begin{tabular}{p{2.5cm}<{\centering}p{4cm}<{\centering}}
         \hline
            coefficient & constraint [$\mathrm{TeV}^{-4}$] \\
         \hline
         $f_{S, 0}/\Lambda^{4}$ & $[-211, 366]$  \\
         $f_{S, 1}/\Lambda^{4}$ & $[-207, 364]$ \\
         $f_{S, 2}/\Lambda^{4}$ & $[-213, 364]$  \\
         $f_{M, 0}/\Lambda^{4}$ & $[-13.2, 30.4]$ \\
         $f_{M, 1}/\Lambda^{4}$ & $[-36.7, 22.9]$ \\
         $f_{M, 2}/\Lambda^{4}$ & $[-11.8, 13.0]$ \\
         $f_{M, 3}/\Lambda^{4}$ & $[-23.1, 20.6]$ \\
         $f_{M, 4}/\Lambda^{4}$ & $[-26.2, 36.8]$ \\ 
         $f_{M, 5}/\Lambda^{4}$ & $[-22.5, 31.5]$ \\
         $f_{M, 7}/\Lambda^{4}$ & $[-43.3, 69.9]$ \\
         $f_{T, 0}/\Lambda^{4}$ & $[-4.63, 3.28]$ \\
         $f_{T, 1}/\Lambda^{4}$ & $[-4.51, 3.34]$ \\
         $f_{T, 2}/\Lambda^{4}$ & $[-9.38, 5.84]$ \\
         $f_{T, 3}/\Lambda^{4}$ & $[-9.22, 6.00]$ \\
         $f_{T, 4}/\Lambda^{4}$ & $[-14.8, 11.5]$ \\
         $f_{T, 5}/\Lambda^{4}$ & $[-7.01, 5.95]$ \\
         $f_{T, 6}/\Lambda^{4}$ & $[-7.00, 6.06]$ \\
         $f_{T, 7}/\Lambda^{4}$ & $[-14.9 ,11.6]$ \\
         $f_{T, 8}/\Lambda^{4}$ & $[-5.25, 5.04]$ \\
         $f_{T, 9}/\Lambda^{4}$ & $[-10.4, 9.66]$ \\
        \hline
     \end{tabular}
 \end{table}

\begin{table}[h]
     \centering
     \caption{Limits at the 95\% CL on aQGC coefficients for the VBS $\PZ\PZ$ process.}
     \label{tab:table_ZZ_constrain}
     \begin{tabular}{p{2.5cm}<{\centering}p{4cm}<{\centering}}
         \hline 
            coefficient & constraint [$\mathrm{TeV}^{-4}$] \\
         \hline
         $f_{S, 0}/\Lambda^{4}$ & $[-14, 13]$  \\
         $f_{S, 1}/\Lambda^{4}$ & $[-5.8, 6.7]$ \\
         $f_{S, 2}/\Lambda^{4}$ & $[-15, 16]$ \\
         $f_{M, 0}/\Lambda^{4}$ & $[-1.2, 1.1]$ \\
         $f_{M, 1}/\Lambda^{4}$ & $[-3.9, 3.7]$ \\
         $f_{M, 2}/\Lambda^{4}$ & $[-8.0, 8.2]$ \\
         $f_{M, 3}/\Lambda^{4}$ & $[-3.9, 3.8]$ \\
         $f_{M, 4}/\Lambda^{4}$ & $[-3.3, 3.2]$ \\ 
         $f_{M, 5}/\Lambda^{4}$ & $[-2.9, 3.0]$ \\
         $f_{M, 7}/\Lambda^{4}$ & $[-8.3, 8.1]$ \\
         $f_{T, 0}/\Lambda^{4}$ & $[-0.11, 0.082]$ \\
         $f_{T, 1}/\Lambda^{4}$ & $[-0.14, 0.14]$ \\
         $f_{T, 2}/\Lambda^{4}$ & $[-0.27, 0.21]$ \\
         $f_{T, 3}/\Lambda^{4}$ & $[-0.27, 0.22]$ \\
         $f_{T, 4}/\Lambda^{4}$ & $[-1.1, 0.67]$ \\
         $f_{T, 5}/\Lambda^{4}$ & $[-0.32, 0.25]$ \\
         $f_{T, 6}/\Lambda^{4}$ & $[-0.47, 0.42]$ \\
         $f_{T, 7}/\Lambda^{4}$ & $[-0.89, 0.60]$ \\
         $f_{T, 8}/\Lambda^{4}$ & $[-0.47, 0.48]$ \\
         $f_{T, 9}/\Lambda^{4}$ & $[-1.1, 1.0]$ \\
        \hline
     \end{tabular}
 \end{table}

\section{\label{sec:level6} Conclusions and Outlook}
In this paper, we investigate $\PZ\PZ\PZ$ productions at a muon collider with $\sqrt{s}=1\,\PTeV, \mathcal{L}=10\,\abinv$, and VBS $\PZ\PZ$ productions at $\sqrt{s}=10\,\PTeV, \mathcal{L}=10\,\abinv$, together with their sensitivities on aQGC coefficients. For these two processes, we focus on pure-leptonic channel and semi-leptonic channel to find the kinematic features that help to increase the detection potential, such as the distribution of $M_{\ell\ell}$ in pure-leptonic channel and $M_{jj}$ in semi-leptonic channel. We have studied the constraints of all aQGC coefficients at 95\% CL. It turns out that for $\PZ\PZ\PZ$ process, we have supplemented the existing tri-boson aQGC results and for some coefficients such as $f_{T, 0}, f_{T, 1}, f_{T, 2}$ in VBS $\PZ\PZ$ process, our results can give stronger limits than existing results. All this demonstrates a great potential to probe anomalous interactions of gauge bosons at the muon collider, due to it’s higher effective collision energy , cleaner final states and higher probability to emit EW radiation than LHC.

\begin{acknowledgments}
This work is supported in part by the National Natural Science Foundation of China under Grants No. 12150005, No. 12075004, and No. 12061141002, by MOST under grant No. 2018YFA0403900.

\end{acknowledgments}

\nocite{*}

\appendix
\label{sec:appendix}
%%%%%%%%%%%%%%%%%%%%%%%%%%%%%%%%%%%%%%%%%%%%%%%%%%%%%%%%%%%%

\end{document}